\documentstyle[aps,eqsecnum,12pt]{revtex}

\title{
Equation of motion for relativistic compact binaries
\\ with the strong field point particle limit :\\
the second and half post-Newtonian order
}

\author{Yousuke Itoh \footnote{yousuke@astr.tohoku.ac.jp}, 
Toshifumi Futamase\footnote{tof@astr.tohoku.ac.jp}\\
Astronomical Institute, Graduate School of 
Science, Tohoku University \\
Sendai 980-8578, Japan \\ 
and \\
Hideki Asada\footnote{asada@phys.hirosaki-u.ac.jp}, \\
Faculty of Science and Technology, 
Hirosaki University, \\
Hirosaki 036-8561, Japan
}

\begin{document}
\maketitle

\begin{abstract}

 We study the equation of motion appropriate to an inspiralling binary
 star system whose constituent stars
 have strong
internal gravity.
We use the post-Newtonian approximation with the strong field point particle
 limit by which we can introduce into general relativity  
 a notion of
 a point-like particle with strong internal 
gravity 
without using Dirac delta distribution.
 Besides this limit, 
 to deal with strong internal gravity we express the equation of
 motion in surface integral forms and calculate these
 integrals  explicitly.
 As a result we obtain the equation of motion for a binary of
 compact bodies accurate through  the second and half post-Newtonian
 (2.5 PN) order. 
This equation is derived in the harmonic coordinate.
 Our resulting equation perfectly agrees with Damour and Deruelle
 2.5 PN equation of motion. 
 Hence it is found that the 2.5 PN equation of motion
 is applicable to a relativistic compact binary.

\end{abstract}

\begin{flushleft}
PACS Number(s): 04.25.Nx, 
\end{flushleft}

\newcommand{\pa}{\partial}

\section{Introduction} 

A relativistic compact binary system in an inspiralling phase 
is one of the most promising sources
of gravitational waves for detectors 
under construction
such as LIGO\cite{Abramovici}, 
VIRGO\cite{Bradaschia}, GEO600\cite{Hough}, and TAMA300\cite{Kuroda}.
Among those detectors, TAMA300 has made  
rapid progress and has recently started observation
\cite{TAGOSHI}. 

To detect gravitational wave signals directly,  
which are expected to be buried
in noisy data streams,  
the matched filtering technique will be (and, in fact, has been)
used. 
To use this technique, we have to construct reliable wave form
templates, i.e., theoretical prediction of  gravitational wave form  
accurate enough
to match a data stream with templates without missing one phase of
the expected signal.
The equation of motion for the source system is one of the necessary 
ingredients for making such templates.

In an inspiralling phase, 
since the separation of the binary is sufficiently larger
than the radius of the stars, 
the interbody gravity is weak and orbital velocities of stars are small
compared to the velocity of light.
Hence the post-Newtonian approximation
with 
a monopole-truncated object \cite{MTO} description 
is useful.
In fact, various authors have used it to derive an equation of
motion of the system
\cite{DD81a,DD81b,Damour82,Damour83,GK,Kopejkin,GII,Schafer86,JS98,BFP,BFY2Ka,BFY2Kb}.
To obtain sufficiently accurate templates, we have
to derive a highly accurate, say, the 4 PN  equation of motion
\cite{CutlerEtAL,CF94,TN94,TTS,DIS} 
(though, the 2 PN templates \cite{BDIWW,BICWW,Blanchet96} might be
enough to
construct search templates\cite{DP97}).
What we are aiming at is to derive such
an accurate post-Newtonian equation of motion. 
In this paper
we derive the 2.5 PN equation of motion as a step toward this goal.

The 2.5 PN equation of motion for two 
monopole-truncated bodies  
has been derived 
by various authors;
Damour and Deruelle \cite{DD81a,DD81b,Damour82,Damour83}
used the post-Minkowskian
formalism \cite{BDDIM},
Kopejikin and Grishchuk worked on a spherical body 
\cite{GK,Kopejkin},
Sch$\ddot{{\rm a}}$fer \cite{Schafer86}
used the Hamiltonian approach,
and Blanchet, Faye and Ponsot used the 
post-Newtonian approach\cite{BFP}.
Their
resulting equations of motion agree with each other.
Then what can we add?
It seems to us that these derivations have some uncomfortable features.
In the derivation in 
\cite{BDDIM,DD81a,DD81b,Damour82,Damour83,Schafer86,BFP}, 
though armed with the ''Dominant Schwartshild'' condition\cite{Damour83},
one needs to regularize divergences caused by their use of Dirac delta 
distribution.
In the derivation free from any divergences    
\cite{GK,Kopejkin}, 
one applies the post-Newtonian approximation
even
to the inside of a star (recall  
that a neutron star has strong internal gravity). 
Though the derivation in \cite{BFP} is mathematically systematic and
well-devised, one assumes that a 
point mass follows a regularized geodesic equation.
Instead, in our method we can take into account the strong
gravity inside stars by use of the strong-field point particle limit
\cite{Futamase85,Futamase87} and
surface integrals, without assuming a geodesic
equation.

One of our motivations to derive 
an higher post-Newtonian equation of motion  
with our method comes from the recent result on the 3 PN equation
of motion, which   
has been derived by two groups in
independent manners 
(see also \cite{WPY2K}).
Jaranowski and Sch$\ddot{{\rm a}}$fer have used Hamiltonian approach and
succeeded to determine the 3 PN order Hamiltonian \cite{JS98}
except two terms whose coefficients are arbitrary;
the kinetic part \cite{JS98} and the static part \cite{JS99}.
The former has been recently determined
by imposing poincar$\acute{\rm e}$ invariance on their 
equation of motion \cite{DJSY2K}.
Blanchet and Faye have also succeeded to derive
the 3 PN equation of motion \cite{BFY2Ka,BFY2Kb}
using their generalized 
Hadamard partie finie regularization\cite{BFY2Kc,BFY2Kd}.
Though their regularization
method is declared to be well-defined unlike the method used by
Jaranowski and Sch$\ddot{{\rm a}}$fer \cite{JS98}, 
they found that there remains one 
arbitrary constant corresponding to the static part in the work of
Jaranowski and Sch$\ddot{{\rm a}}$fer.
Although the method of Blanchet and Faye seems
mathematically rigorous and well-devised, what is uncomfortable, besides
the above arbitrary constant, comes from the fact that  
they rely on some ''naturality'' : They assume that a mass
point follows a regularized geodesic equation, but
 the assumption should be verified. 
Hence, it is interesting  to derive the 3 PN equation
of motion using yet another method.
Besides this,  It is important to derive the equation of motion with
various methods to confirm
the equation, which appears through long calculations.

In our previous paper (\cite{IFAY2K}, referred as paper I henceforth),
we have derived
the 1 PN equation of motion using the post-Newtonian approximation
with the strong filed point particle limit
\cite{Futamase85,Futamase87} and
the general equation of motion expressed in surface integrals
\cite{IFAY2K} as in the Einstein-Infeld-Hoffmann approach \cite{EIH}.
The spin-orbit and quadrupole-orbit coupling forces which appear at
the 2 PN order in our ordering have been also derived.
Following 
the initial value formulation on the field by
Schutz and Futamase\cite{Schutz80,FS83,Futamase83,Schutz85},
in this paper
we shall derive the 2.5 PN equation of motion.

This paper is organized as follows.
In Sec. II, we will  review  the post-Newtonian approximation
and the strong field point particle limit.
In Sec. III we shall show how to solve
the Einstein equation. There we shall define 
multipole moments
of stars, and introduce a general form of the equation of motion.
It is entirely expressed in surface integral form. 
{}From Sec. IV to Sec. VII, we shall derive the equation of motion
for two compact bodies accurate though the 2.5 PN order, the order
at which the radiation reaction effect first appears.
It will be found that 
the temporal component of the equation of
motion and its (functional) solution play an important role in
our derivation.
The section VIII is devoted to conclusion and discussion.  
Useful formulas and explanations on subtleties of our
method will be given in appendices.

Throughout we use units where $c=1=G$. $\vec x$ denotes
Euclidean three vector. We raise or lower its index 
with Kronecker delta.

\section{post-Newtonian approximation with strong field point
particle limit}
\label{PNAwithSFPPL}

In this section we
shall review the strong field point particle limit 
associated with 
the post-Newtonian approximation.
Also scalings of initial data for matter and a body zone will be explained. 
These scalings enable us to incorporate a self-gravitating point-like 
particle into post-Newtonian approximation and ensure that tidal
effect is the lowest order effect which causes an internal motion of the 
stars except their spinning motion. See \cite{Futamase85,Futamase87} 
for detail.

In the inspiralling phase, the post-Newtonian approximation with a
point particle description is suitable to derive the equation of motion.
In the post-Newtonian approximation we assume the balance
between 
Newtonian gravitational force and  centrifugal force, i.e.,
$m/L^2 \sim v_{orb}^2/L$, where m, L, and $v_{orb}$ are typical scales of
a mass of a star, the separation of
the binary, and an orbital velocity of the star.
Then a natural time scale $\tau$, which we call Newtonian 
dynamical time, is $\tau = \epsilon t$\cite{FS83},  
where $\epsilon$ is a non-dimensional parameter and represents 
the smallness of the orbital velocity;
\begin{eqnarray}
&&
v^i_{orb} = \frac{d x^i}{d t} = \epsilon \frac{d x^i}{d \tau},
\end{eqnarray}
and we set $dx^i/d\tau = O(\epsilon^0)$.
Therefore we take $\epsilon$ as the post-Newtonian expansion parameter.
Then, let  $\epsilon$ go to zero with $\tau$ remaining constant.
The meaning of $\tau$ is that 
two events at the same $(\tau, x^i)$ but having different $\epsilon$
are in the nearly same phase in the orbit\cite{FS83}.
Then from the Newtonian
force balance, $m \sim \epsilon^2$ when we keep the orbital separation
constant. 

Now we achieve point particle limit by letting the radius of the 
star, say, $R$, shrink
at the same rate as the mass of the star;
$m/R = O(\epsilon^0) = $ constant while $R = O(\epsilon^2) \rightarrow 0$ and
$m  = O(\epsilon^2) \rightarrow 0$ as $\epsilon \rightarrow 0$. Because of  
finite $m/R$, we say this point particle has strong internal gravity.  
This scaling implies that the density changes proportional
to $\epsilon^{-4}$ (in the $(t,x^i)$ coordinate). 
The scalings of $R$ and $m$ motivate
us to define the body zone of the star A (A=1,2),
$B_A =\{x^i | |\vec{x} - \vec{z}_A(\tau)| < \epsilon R_A \}$
and a body zone coordinate of the star A, 
$\alpha_A^{\underline{i}} = \epsilon^{-2} (x^i - z_A^i(\tau))$.
$z_A^i(\tau)$ is
a representative point of the star A, e.g., the center of the mass
of the star A. $R_A$ is an arbitrary 
length scale (much smaller than $L$ and is not the radius of the star)
and constant (, i.e., $d R_A/d\tau =0)$. With the body zone coordinate,
the star does not shrink, while the boundary of the body zone
goes to infinity. Then it is appropriate to define
star's characteristic quantities such as a mass, a spin, and so on with
the body zone coordinate. On the other hand the body zone serves us  
with a surface $\pa B_A$, through which gravitational energy momentum flux 
flows and  in turn it amounts to gravitational force exerting on the
star A. 
Because the body zone boundary $\pa B_A$ is far away from the surface
of the star A, we can evaluate the gravitational energy momentum flux
at $\pa B_A$ 
with the post-Newtonian gravitational field. In fact
we shall express
our equation of motion in terms of integrals over $\pa B_A$ and
be able to evaluate them explicitly.

Let us move on to the scaling of the matter variables.
As in the paper I,
we consider an initial value problem
to solve the Einstein equation. We take 
two nearly stationary solutions
of the exact Einstein equation as the
initial data for the matter variables (and gravitational field).  
Then we  assume that these solutions
have the following scalings; the density changes proportional to
$\epsilon^{-4}$, the velocity of spinning motion is $O(\epsilon)$.   
The slow spinning motion assumption is not crucial: In fact,
it is straightforward to
incorporate a rapidly spinning compact body into our formalism.
The scaling of the density suggests that the natural dynamical time 
(free fall time) $\eta$ inside the star may be $\eta = \epsilon^{-2}t$. Then
if we can not assume the nearly stationary condition on the stars,
it is difficult to use the post-Newtonian approximation
\cite{Futamase85,Futamase87}.

{}From these initial data we have the following scalings of the star A's
stress energy tensor components in the body zone coordinate,
$T_A^{\mu\nu}$;
$T_A^{\tau\tau} = O(\epsilon^{-2})$,
$T_A^{\tau \underline{i}} = O(\epsilon^{-4})$,    
$T_A^{\underline{i}\underline{j}} = O(\epsilon^{-8})$.
Here the underlined indices mean that for any tensor $A^i$,
$A^{\underline i} = \epsilon^{-2} A^i$.
In the paper I,
we have transformed $T_N^{\mu\nu}$,
the components of the stress energy tensor of the matter in the
near zone coordinate, to  $T_A^{\mu\nu}$ using the transformation
of
Fermi normal coordinate at the 1 PN order.
It is difficult, however, to
construct the Fermi normal coordinate at an higher post-Newtonian order. 
Therefore we shall not use it.
We simply assume that for $T_N^{\mu\nu}$ (or rather
$\Lambda_N^{\mu\nu}$, see Eq. (\ref{DefOfLambda})),

\begin{eqnarray}  
&&
T_N^{\tau\tau} = O(\epsilon^{-2}),
\label{ScalingTNtt} \\
&&
T_N^{\tau \underline{i}} = O(\epsilon^{-4}),
\label{ScalingTNti} \\ 
&&
T_N^{\underline{i}\underline{j}} = O(\epsilon^{-8}),
\label{ScalingTNij} 
\end{eqnarray}
as their leading scalings. We have found that these scalings
are enough to derive the 2.5 PN equation of motion for two 
compact bodies.

Henceforth we call the coordinate $(\tau,x^i)$ the near zone coordinate.

\section{Field equation and the general form of the equation of motion}

\subsection{Field equation}

As discussed in the previous section,
we shall express our equation of motion in terms of
surface integrals over the body zone boundary where it is assumed
that 
the metric slightly deviates from the flat metric
$\eta^{\mu\nu} = {\rm diag}{(-\epsilon^2,1,1,1)}$
(in the near zone coordinate).
Thus we define a deviation
field $h^{\mu\nu}$ as
\begin{eqnarray}
&&
h^{\mu\nu} =  \eta^{\mu\nu} - \sqrt{-g}g^{\mu\nu},
\end{eqnarray}
where $g$ is the determinant of the metric. Our $h^{\mu\nu}$
differs from the corresponding field in \cite{BFP} in a sign.

Now we choose the harmonic coordinate condition on the metric
\begin{equation}
 h^{\mu\nu}{}_{,\nu}=0 , 
\end{equation}
where the comma denotes the partial derivative. 
Then, we recast the Einstein equation into the relaxed 
form, 
\begin{equation}
\Box h^{\mu\nu} = -16\pi \Lambda^{\mu\nu} , 
\end{equation}
where $\Box = \eta^{\mu\nu}\pa_{\mu}\pa_\nu$
is the flat d'Alembertian and   
\begin{eqnarray}
&&\Lambda^{\mu\nu} = \Theta^{\mu\nu}
+\chi^{\mu\nu\alpha\beta}{}_{,\alpha\beta} ,
\label{DefOfLambda} \\ 
&&\Theta^{\mu\nu} = (-g) (T^{\mu\nu}+t^{\mu\nu}_{LL}) , \\
&&\chi^{\mu\nu\alpha\beta} = \frac{1}{16\pi} 
(h^{\alpha\nu}h^{\beta\mu}
-h^{\alpha\beta}h^{\mu\nu}) . 
\end{eqnarray}
Here, $T^{\mu\nu}$ and $t^{\mu\nu}_{LL}$ denote the stress-energy tensor 
of the stars 
and the Landau-Lifshitz pseudo-tensor \cite{LL}.
$\chi^{\mu\nu\alpha\beta}$ originates from our use of the flat d'Alembertian 
instead of 
the curved space d'Alembertian. 
In consistency with the harmonic condition, the
conservation law is expressed as
\begin{equation}
\Lambda^{\mu\nu}{}_{,\nu}=0 . 
\label{conservation}
\end{equation}

Now we rewrite the relaxed Einstein equation into an integral form,  
\begin{equation}
h^{\mu\nu}=4 \int_{C(\tau, x^k; \epsilon)} d^3y 
\frac{\Lambda^{\mu\nu}(\tau-\epsilon|\vec x-\vec y|, y^k; \epsilon)} 
{|\vec x-\vec y|} ,  
\label{IntegratedREE}
\end{equation}
where $C(\tau, x^k;\epsilon)$ means the past light cone emanating 
{}from the event $(\tau, x^k)$ (about the dependence of $C$ on
$\epsilon$, see \cite{FS83}).  
 We ignored the homogeneous solution, which corresponds to
the initial data for the field, 
for simplicity. Even if we take random initial data for the field 
\cite{Schutz80} supposed to be of the 1 PN order\cite{Futamase83},
they are irrelevant to the dynamics of the binary system up to
the radiation reaction order 
inclusively
\cite{Futamase83}. 

We shall solve the relaxed Einstein equation as follows.
First, we make retarded expansion of Eq. (\ref{IntegratedREE}) 
and change the integral region
$C$ to $N$
(, $N$ denoting near zone), 
a $\tau =$ constant spatial
hypersurface 
(note that such retarded expansion is understood to be
formal, since it gives divergent integrals. Because such divergent
integrals do not appear up to the 2.5 PN order and  
we are concerned with the 2.5 PN equation of motion in this paper,  
we use formal retarded expansion. See below).
\begin{equation}
h^{\mu\nu}=4
\sum_{n=0} \frac{(-\epsilon)^n}{n!}\left(\frac{\pa}{\pa \tau}\right)^n
\int_{N} d^3y 
|\vec x-\vec y|^{n-1}
\Lambda_N^{\mu\nu}(\tau, y^k; \epsilon).
\label{RetardedEIREE}
\end{equation}
Here we added a subscript $N$ to $\Lambda^{\mu\nu}$ to clarify
that $\Lambda^{\mu\nu}$ here are components in the near zone
coordinate. Simple retarded expansion gives divergent integrals
at the 3 PN $ $order.
In this paper we take the near zone as a sphere centered at
some fixed point and enclosing the binary system. The radius
of this sphere is set to be
${\cal R}/\epsilon$, where ${\cal R}$ is arbitrary
but much larger  than the size of the binary. Then
we find that up to the 2.5 PN order, 
the terms dependent on ${\cal R}$ take
forms such as ${\cal R}^{-n}$, $\pa_{\tau}{\rm ln}{\cal R}$, 
$\pa_{i} {\rm ln}{\cal R}$,
$\pa_{\tau}{\cal R}^n$, or $\pa_{i}{\cal R}^n$ ($n \ge 1$), where
$\pa_{\mu} = \pa/\pa x^{\mu}$.    
Hence we simply let ${\cal R}$ go to infinity.

Second we split
the integration into two parts; contribution from the body zone
$B_A$, 
and from elsewhere, $N/B$. Schematically we evaluate
the following type integration (we omit indices of the field),  
\begin{eqnarray}
&&
h = h_B + h_{N/B}, \\
&&
h_B = 
\epsilon^6 \sum_{A=1,2}\int_{B_A}
 d^3\alpha_A 
\frac{f(\tau,\vec z_A + \epsilon^2\vec{\alpha}_A)}
{|\vec r_A-\epsilon^2\vec{\alpha}_A|^{1-n}} ,  
\\
 &&
h_{N/B} = \int_{N/B}  d^3y 
\frac{f(\tau,\vec{y})}
{|\vec x-\vec y|^{1-n}},
\label{NBContributionSC}
 \end{eqnarray}
where $\vec r_A = \vec x - \vec z_A$.
We shall deal with these two contributions successively.

\subsubsection{Body zone contribution}

As for the body zone contribution,  
we make  multipole expansion with care over 
the scaling of the integrand, i.e., 
$\Lambda^{\mu\nu}$ in the body zone. 
For example, $n=0$ part in  Eq. (\ref{RetardedEIREE}),
$h_{B n=0}^{\mu\nu}$, gives
\begin{eqnarray}
&&
h^{\tau\tau}_{B n=0} = 4 \epsilon^4 \sum_{A=1,2}
\left(\frac{P_A^{\tau}}{r_A} + \epsilon^2 \frac{D_A^k r^k_A}{r_A^3} +
 \epsilon^4 \frac{3 \hat{I}_A^{kl} r^k_A r^l_A}{2 r_A^5} \right)
 + O(\epsilon^{10}),
\label{hBtt} \\ 
&&
h^{\tau i}_{B n=0} = 4 \epsilon^4 \sum_{A=1,2}
\left(\frac{P_A^{i}}{r_A} + \epsilon^2 \frac{J_A^{ki} r^k_A}{r_A^3} 
 \right)
 + O(\epsilon^8),
\label{hBti} \\ 
&&
h^{ij}_{B n=0} = 4 \epsilon^2 \sum_{A=1,2}
\left(\frac{Z_A^{ij}}{r_A} + \epsilon^2 \frac{Z_A^{kij} r^k_A}{r_A^3} +
 \epsilon^4 \frac{3 \hat{Z}_A^{klij} r^k_A r^l_A}{2 r_A^5} \right)
 + O(\epsilon^8).
 \label{hBij}
\end{eqnarray}
Here the quantity with a  
hat 
is symmetric and tracefree on its dummy indices, 
and $ r_A = |\vec{r}_A|$.
To derive the 2.5 PN equation of motion,   
$h^{\tau\tau}$ up to $O(\epsilon^9)$ and 
$h^{\tau i}$ and $h^{i j}$ up to $O(\epsilon^7)$ 
are needed.  

In the above equations we defined multipole moments as 
\begin{eqnarray}
&&
I_A^{L} = \epsilon^2 \int_{B_A}
d^3\alpha_A \Lambda^{\tau \tau}_N \alpha_A^{\underline L}, \\   
&&
J_A^{Li} = \epsilon^4 \int_{B_A}
d^3\alpha_A \Lambda^{\tau \underline i}_N \alpha_A^{\underline L},  \\
&&
 Z_A^{Lij} = \epsilon^8 \int_{B_A}
d^3\alpha_A \Lambda^{\underline{i}\underline{j}}_N \alpha_A^{\underline L},  
\end{eqnarray}
where  we introduced  multi index notations 
$L = i_1 i_2 \cdot\cdot\cdot i_l$  and
$\alpha_A^{\underline L} =
\alpha_A^{\underline i_1}\alpha_A^{\underline i_2}
\cdot\cdot\cdot\alpha_A^{\underline i_l}$.  
Then 
$P_A^{\tau} = I^{0}_A$, $D_A^{i} = I^{i}_A$,
$P_A^{i} = J^{i}_A $.
We simply call $P^{\mu}_A$ the four momentum of the star A,
$P^{i}_A$ the momentum, and $P^{\tau}_A$ the energy. 
Also we call $D^{k}_A$ the dipole moment of the star A, and
$I^{kl}_A$ the quadrupole moment of the star A.

Then we transform these moments into more convenient forms.  
By the conservation law Eq. (\ref{conservation}), we have
\begin{eqnarray}
&&
\Lambda_N^{\tau i} =  (\Lambda_N^{\tau\tau} y_A^i)_{,\tau} +
 (\Lambda_N^{\tau j} y_A^i)_{,j} + v_A^i \Lambda_N^{\tau \tau}, \\
&&
\Lambda_N^{i j} =  (\Lambda_N^{\tau (i} y_A^{j)})_{,\tau} +
 (\Lambda_N^{k (i} y_A^{j)})_{,k} + v_A^{(i} \Lambda_N^{j) \tau},
\end{eqnarray} 
where $v_A^i = \dot{z}^i_A$, 
an overdot 
denotes $\tau$ time
derivative, and $\vec y_A = \vec y - \vec z_A$. 
Using these equations and noticing that the body zone remains unchanged
(in the near zone coordinate), i.e., $\dot R_A = 0$, we have 
\begin{eqnarray}
&&
P^{i}_A = P^{\tau}_A v^i_A + Q_A^i +
\epsilon^2 \frac{d D_A^i}{d\tau},    
\label{MomVelRelation} \\ 
&&
J^{ij}_A =
\frac{1}{2}
\left(M_A^{ij} + \epsilon^2 \frac{d I_A^{ij}}{d\tau}\right) 
+ v_A^{(i} D_A^{j)} + \frac{1}{2}\epsilon^{-2} Q_A^{ij},
\label{JijToMij}
\\
Z^{ij}_A &=& \epsilon^2 P^{\tau}_A v_A^i v_A^j + 
 \frac{1}{2}\epsilon^6\frac{d^2 I_A^{ij}}{d\tau^2} +
2\epsilon^4 v_A^{(i}\frac{d D_A^{j)}}{d\tau} +
\epsilon^4 \frac{d v_A^{(i}}{d\tau}D_A^{j)}  \nonumber \\
\mbox{} &+& 
\epsilon^2 Q_A^{(i}v_A^{j)} + \epsilon^2 R_A^{(ij)} +
\frac{1}{2}\epsilon^2\frac{d Q_A^{ij}}{d\tau},
\label{StrVelRelation}
\\
&&
Z^{kij}_A = \frac{3}{2}A_A^{kij} - A_A^{(ij)k},
\label{ZkijToAkij}
\end{eqnarray}
where 
\begin{eqnarray}
&&
M_A^{ij} = 2\epsilon^4\int_{B_A}d^3\alpha_A
\alpha_A^{[\underline i}\Lambda_N^{\underline j]\tau} , \\
&&
Q_A^{Li} = \epsilon^{-4}
 \oint_{\pa B_A} dS_k
 \left(\Lambda^{\tau k}_N  - v_A^k 
\Lambda^{\tau\tau}_N \right) y_A^L y_A^i
\label{QL} , \\ 
&&
R_A^{Lij} = \epsilon^{-4}
 \oint_{\pa B_A} dS_k
 \left(\Lambda^{k i}_N  - v_A^k 
\Lambda^{\tau i}_N \right) y_A^L y_A^j,
\label{RL} 
\end{eqnarray}
and
\begin{eqnarray}
&&
A_A^{kij} = \epsilon^2 J_A^{k(i}v_A^{j)} + \epsilon^2 v_A^k J_A^{(ij)}
+R_A^{k(ij)} + \epsilon^4 \frac{dJ_A^{k(ij)}}{d\tau}.
\end{eqnarray}
$[$ $]$ and $($ $)$ attached to indices
denote anti-symmetrization and symmetrization.
$M_A^{ij}$ is the spin of the star A and  Eq. (\ref{MomVelRelation})
is the momentum-velocity relation. 
In general, we have 
\begin{eqnarray}
&&
J_A^{Li} = J_A^{(Li)} +  \frac{2l}{l+1}J_A^{(L-1[i_l)i]}, \\
&&
Z_A^{Lij} = \frac{1}{2}
\left[Z_A^{(Li)j} + \frac{2 l}{l+1}Z_A^{(L-1[i_l)i]j} +
Z_A^{(Lj)i} + \frac{2 l}{l+1}Z_A^{(L-1[i_l)j]i} \right],
\end{eqnarray}
and 
\begin{eqnarray}
&&
J_A^{(Li)} = \frac{1}{l+1}\epsilon^2\frac{d I_A^{Li}}{d\tau}
+ v_A^{(i}I_A^{L)} + \frac{1}{l+1}\epsilon^{-2l}Q_A^{Li}, 
 \label{JLiToILi}
\\
&&
Z_A^{(Li)j} + Z_A^{(Lj)i} =
\epsilon^2 v_A^{(i}J_A^{L)j} +
\epsilon^2 v_A^{(j}J_A^{L)i} +
\frac{2}{l+1}\epsilon^4\frac{d J_A^{L(ij)}}{d\tau}
+ \epsilon^{-2l + 2}R_A^{L(ij)}, 
\label{ZLijToJLij} 
\end{eqnarray}
where $l$ is the number of indices in the multi index L. 

Now, from the above equations, especially
 Eq. (\ref{StrVelRelation}), we find that the body zone
contribution, 
$h^{\mu\nu}_{B n=0}$, are of order $O(\epsilon^4)$. Note that if we
can not or do not assume the (nearly) stationarity of the initial data
for the stars,
then, instead of Eq. (\ref{StrVelRelation}) we have  
$$
Z^{ij}_A = \epsilon^2 P^{\tau}_A v_A^i v_A^j + 
 \frac{1}{2}\frac{d^2 I_A^{ij}}{d\eta^2} + \cdot\cdot\cdot,
$$
where we used the dynamical time $\eta$ (see Sec \ref{PNAwithSFPPL}).
In this case the lowest metric differs from the Newtonian form.
{}From our (nearly) stationary assumption the remaining motion inside
a star, except the spinning motion,  is caused only by the tidal
effect by the companion star and from  Eq. (\ref{StrVelRelation}),
it appears at the 3 PN order\cite{Futamase87}.

To obtain the lowest order $h^{\mu\nu}_{B n=0}$,
we have to
evaluate the surface integrals $Q_A^{Li}, R_A^{Lij}$.
Generally, in  $h_{B n=0}^{\mu\nu}$
the moments $J_A^{Li}$ and $Z_A^{Lij}$ appear formally at the order
$\epsilon^{2 l + 4}$ and $\epsilon^{2 l + 2}$. Thus $Q^{Li}_A$ and 
$R_A^{Lij}$ appear as 
$$
h^{\tau i}_{B n=0} \sim
\cdot\cdot\cdot + 
\epsilon^4 \frac{r_A^L \hat Q^{Li}_A}{r_{A}^{2l+1}} + \cdot\cdot\cdot,
$$
$$
h^{ij}_{B n=0} \sim
\cdot\cdot\cdot + 
\epsilon^4 \frac{r_A^L \hat R^{Lij}_A}{r_{A}^{2l+1}} + \cdot\cdot\cdot, 
$$
where we omitted irrelevant terms and numerical coefficients. 
Thus one may expect that $Q_A^{Li}$ and $R_A^{Lij}$ appear at the
order 
$\epsilon^4$ for any L and we have to calculate an infinite number
of moments. This is not the case. 
We address this problem in the appendix \ref{RLijandQLi}.
The important thing here is
that $\epsilon^4 Q_A^{Li}$ and $\epsilon^4 R_A^{Lij}$ are
at most $O(\epsilon^4)$ in $h^{\mu\nu}_{B n=0}$.

Finally, since the order of
$h_{B n}^{\mu\nu} (n \ge 1)$ is higher than that of
$h_{B n=0}^{\mu\nu}$, $h_B^{\mu\nu} = O(\epsilon^4)$.

\subsubsection{$N/B$ contribution}

About the $N/B$ contribution, 
since the integrand 
$\Lambda_N^{\mu\nu} = -gt_{LL}^{\mu\nu} +
\chi^{\mu\nu\alpha\beta}\mbox{}_{,\alpha\beta}$ is at most
quadratic in the small deviation field $h^{\mu\nu}$,
we make the post-Newtonian expansion in the integrand. 
Then, basically, with the help of a potential $g(\vec{x})$
which satisfies $\Delta g(\vec x) = f(\vec{x})$,
$\Delta$ denoting Laplacian, we have for each integral (, e.g.,
$n=0$ term in Eq. (\ref{NBContributionSC}))    
\begin{eqnarray}
\int_{N/B} d^3y \frac{f(\vec{y})}{|\vec{x}-\vec{y}|}  &=&
- 4 \pi g(\vec{x}) + \oint_{\pa(N/B)}dS_k
\left[\frac{1}{|\vec{x}-\vec{y}|}
 \frac{\pa g(\vec{y})}{\pa y^k} - g(\vec{y})\frac{\pa}{\pa y^k}
 \left(\frac{1}{|\vec{x} - \vec{y}|}\right) \right].  
\label{NBcontribution}
\end{eqnarray} 
The last surface integral really contributes to the 2.5 PN gravitational
field. For $n \ge 1$ terms in Eq. (\ref{NBContributionSC}), we use
potentials many times to convert all the volume integrals into surface
integrals and ``$- 4 \pi g(\vec x)$`` terms.

Now the lowest order integrands can be evaluated   
with the body zone contribution $h_B^{\mu\nu}$, and since $h_B^{\mu\nu}$ 
is $O(\epsilon^4)$, we find (see appendix \ref{pNELLPT})
\begin{eqnarray}
&&
\Lambda_N^{\tau\tau} = O(\epsilon^6),
\label{tLLttlowest}\\  
&&
\Lambda_N^{\tau i} = O(\epsilon^6),
\label{tLLtilowest}\\
&&
\Lambda_N^{ij} = (-g)t_{LL}^{ij} + O(\epsilon^8) = 
\epsilon^4 \frac{1}{64 \pi}
\left(\delta^i\mbox{}_k\delta^j\mbox{}_l
 - \frac{1}{2}\delta^{ij}\delta_{kl}\right)
\mbox{}_4h_B^{\tau\tau ,k}\mbox{}_4h_B^{\tau\tau ,l} + O(\epsilon^5).
\label{tLLijlowest}
\end{eqnarray}
where we expanded $h_B^{\mu\nu}$ in an $\epsilon$ series;
$$
h_B^{\mu\nu} = \sum_{n=0}\epsilon^{4+n}\mbox{}_nh_B^{\mu\nu}.
$$
Similarly, in the following we expand $h^{\mu\nu}$ in an $\epsilon$
series. 
{}From these equations we find that
the deviation field in N/B, $h^{\mu\nu}$, is $O(\epsilon^4)$ 
(It should be noted that in the body zone $h^{\mu\nu}$ is assumed to be of
order unity and within our method 
we can not calculate $h^{\mu\nu}$ there explicitly.
To obtain $h^{\mu\nu}$ in the body zone,
we have to solve the internal problem).  

The Landau-Lifshitz pseudo-Tensor
expanded with $h^{\mu\nu}$ is listed in the appendix \ref{pNELLPT}.

\subsection{General Form of the Equation of Motion}

{}From the definition of the four momentum 
\begin{eqnarray}
&&
P^{\mu}_A(\tau)= \epsilon^2 \int_{B_A} d^3\alpha_A \Lambda^{\tau \mu}_N, 
\label{DefOfMomentum} 
\end{eqnarray}
and the conservation law, Eq. (\ref{conservation}), we have the 
evolution equation for the four momentum; 
\begin{equation}
\frac{dP_A^{\mu}}{d\tau} = -\epsilon^{-4}
 \oint_{\pa B_A} dS_k \Lambda^{k\mu}_N
+\epsilon^{-4}
v_A^k \oint_{\pa B_A} dS_k \Lambda^{\tau\mu}_N. 
\label{EvolOfFourMom}
\end{equation}
Here we used the fact that the size and the shape of the
body zone are defined to be fixed (in the near zone coordinate).

Inserting the momentum-velocity relation, Eq. (\ref{MomVelRelation}), 
into the spatial component of  Eq. (\ref{EvolOfFourMom}), we obtain
the general form of the equation of motion for the star A;   
\begin{eqnarray}
P_A^{\tau}\frac{dv_A^i}{d\tau} =&&
 -\epsilon^{-4}
 \oint_{\pa B_A} dS_k \Lambda^{ki}_N
+ \epsilon^{-4}
v_A^k \oint_{\pa B_A} dS_k \Lambda^{\tau i}_N 
\nonumber\\
&&
 +\epsilon^{-4}
 v_A^i \left( \oint_{\pa B_A} dS_k \Lambda^{k\tau}_N
-v_A^k \oint_{\pa B_A} dS_k \Lambda^{\tau\tau}_N \right)
\nonumber\\
&&-\frac{dQ_A^i}{d\tau}  - \epsilon^2 \frac{d^2 D_A^i}{d \tau^2}.  
\label{generaleom}
\end{eqnarray} 

All the right hand side terms in Eq. (\ref{generaleom}) except the
dipole moment are surface integrals. 
We can specify the
value of $D_A^i$ freely to determine the representative point
$z_A^i(\tau)$ of the star A.
In this paper we take $D_A^i = 0$ and simply call $z_A^i$ the
center of the mass of the star A. Note that to obtain the 
spin-orbit coupling force in the same form as the previous works
\cite{Damour82,TH,Kidder},
we have to make another choice for $z_A^i$ (paper I).
For completeness we show
in the appendix \ref{SOcoupling} 
the spin-orbit coupling
force on the condition that $D_A^i =0$. 
Henceforth we shall restrict
our attention on the mass monopoles.

In Eq. (\ref{generaleom}), $P_A^{\tau}$
rather than the mass of the star A appears. Hence
we have
to derive a relation between the mass and $P_A^{\tau}$.
We shall achieve this by solving the temporal component of
the evolution equation (\ref{EvolOfFourMom}) functionally.

Then since all the equations are expressed with surface integrals
except $D_A^i$ to be specified, 
we can derive the
equation of motion
for the strongly self-gravitating star
using the post-Newtonian approximation 
at least before the
order where we have to be concerned with the internal problem.

\subsection{On the Arbitrary Constant $R_A$}
\label{Arbitrariness}

Since we introduce the body zone by hand,
the arbitrary defined body zone size $R_A$
seems to appear in the metric, the multipole moments of the stars,  
and the equation of motion.
More specifically $R_A$ appears in them because of     
1) the splitting of the deviation field into two parts  
(, i.e., $B$ and $N/B$ contributions),
the definition of the moments, and  
2) the surface integrals that we evaluate to derive the
equation of motion.

\subsubsection{$R_A$ Dependence Of The Field }

$B$ and $N/B$ contributions 
to the field 
depend on 
the body zone boundary, $\epsilon R_A$.
But $h^{\mu\nu}$ itself
does not depend on $\epsilon R_A$.
Thus it is natural to expect that there are renormalized multipole moments 
which are independent of $R_A$ since we use non-singular matter
sources. One possible practical obstacle for this expectation might 
be $\log(\epsilon R_A)$ dependence of multipole moments.  We have checked that 
at least up to the 2.5 PN order, there is no such log-term.

Though we use the same symbol for the moments henceforth as before for
notational simplicity, it should be  understood that they are the
renormalized ones (we use the symbol ''$P_A^{\mu}$''
for the renormalized $P_A^{\mu}$).

\subsubsection{$R_A$ Dependence Of The Equation Of Motion}

Since we make integrations over the body zone boundary,
in general the resulting equation of motion seems to depend
on the size of the body zone boundary, $\epsilon R_A$.  
But it does not depend on $R_A$.

In the derivation of Eq. (\ref{generaleom}), 
if we did not use the conservation law (\ref{conservation})  
until the final step, we have  
\begin{eqnarray}
P_A^{\tau}\frac{d v_A^i}{d\tau} +&&
  \epsilon^{-4}
 \oint_{\pa B_A} dS_k \Lambda^{ki}_N
-  \epsilon^{-4}
v_A^k \oint_{\pa B_A} dS_k \Lambda^{\tau i}_N 
\nonumber\\
&&
 - \epsilon^{-4}
 v_A^i \left( \oint_{\pa B_A} dS_k \Lambda^{k\tau}_N
- v_A^k \oint_{\pa B_A} dS_k \Lambda^{\tau\tau}_N \right)
\nonumber\\
&&+\frac{d Q_A^i}{d\tau}  + \epsilon^2 \frac{d^2 D_A^i}{d \tau^2}   
\nonumber\\
&&=
\epsilon^{-4}\int_{B_A}d^3y\Lambda_N^{i \nu}\mbox{}_{,\nu}
- \epsilon^{-4}v_A^i\int_{B_A}d^3y\Lambda_N^{\tau \nu}\mbox{}_{,\nu}
+\epsilon^{-4}\frac{d}{d\tau}\left(\int_{B_A}d^3y
\Lambda_N^{\tau \nu}\mbox{}_{,\nu}y_A^i\right). 
\end{eqnarray}
Now the conservation law is satisfied for whatever value we take as $R_A$,
then the right hand side of the above equation is zero for any $R_A$.
Hence the equation of motion Eq. (\ref{generaleom}) does not
depend on $R_A$ (Similar argument can be found in \cite{EIH}).

Along the same line, 
the momentum-velocity relation (\ref{MomVelRelation})
does not depend on $R_A$.

Up to the 1 PN order in paper I,  
we have explicitly shown the irrelevance of the equation of motion
to $R_A$ by checking the cancellation among
the $R_A$ dependent
terms.

\section{Newtonian and the 1 PN equation of motion}
\label{NandFPNEOM}

In this section we shall
list the Newtonian and the 1 PN equation of motion
for later convenience.
As for the details of the derivation in our method, see
the paper I.

Let us review how to derive the field and the
equation of motion order by order.

First of all, from Eqs. (\ref{tLLttlowest}), (\ref{tLLtilowest}),
and the 
time component of Eq. (\ref{EvolOfFourMom}), we have    
\begin{eqnarray}
&&
 \frac{d P_A^{\tau}}{d\tau} = O(\epsilon^2).
\label{Eq4.1}
\end{eqnarray}
Then we define the mass of the star A as  
the integrating constant of this equation;   
\begin{eqnarray}
&& 
m_A = \lim_{\epsilon \to 0}P_A^{\tau}.  
\label{DefOfMass}
\end{eqnarray}
$m_A$ is the ADM mass that the star A had if A were isolated 
(we took $\epsilon$ zero limit in Eq. (\ref{DefOfMass}) 
to ensure that the mass defined above  does not include the effect of the
companion star and the orbital motion of the star itself. By this limit 
we ensure that this mass is the integrating constant of Eq. (\ref{Eq4.1}). 
Some subtleties
about this definition are discussed in the appendix
\ref{SubtltyOfMass}).
By definition $m_A$ is constant.
Then we obtain the lowest order $h^{\tau\tau}$;
\begin{eqnarray}
&&
\mbox{}_4h^{\tau\tau} = 4 \sum_{A=1,2}
\frac{m_A}{r_A}.  
\end{eqnarray}

Second, from Eq. (\ref{QL}) with  
Eqs. (\ref{tLLttlowest}) and (\ref{tLLtilowest}) we obtain
$Q_A^i = 0$ at  
the lowest order.
Thus we have the Newtonian momentum-velocity
relation $P^{i}_A = m_A v_A^i + O(\epsilon^2)$ from Eq.
(\ref{MomVelRelation}) (we set $D^i_A = 0$). 
Then from Eq. (\ref{generaleom}), we can derive the Newtonian
equation of motion. This completes the Newtonian order calculations.

Next, at the 1 PN order, we need $\mbox{}_6h^{\tau\tau}$,
$\mbox{}_4h^{\mu\nu}$.
The $n=1$ term in the retardation expansion series of $h^{\tau\tau}$, 
Eq. (\ref{RetardedEIREE}), gives no contribution at the 1 PN
order by the constancy of the mass $m_A$, i.e.,
$\mbox{}_5h^{\tau\tau} =0$.

Now we  obtain $\mbox{}_4h^{\tau i}$ from
the Newtonian momentum-velocity relation.
$$
\mbox{}_4h^{\tau i} = 4  \sum_{A=1,2}
\frac{m_A v_A^i}{r_A}.  
$$

{}From $\mbox{}_4h^{\tau\nu}$ and Eqs. (\ref{tLLtt6}) and (\ref{tLLti6})
we can evaluate surface integrals in the evolution equation for $P_A^{\tau}$
at the 1 PN order.
The result is (for the star 1) 
\begin{equation}
\frac{dP_1^{\tau}}{d\tau}=\epsilon^2 m_1 \frac{d}{d\tau} 
\left( \frac12 v_1^2+\frac{3m_2}{r_{12}} \right) + O(\epsilon^3), 
\end{equation}
where $\vec{r}_{12} = \vec{z}_1 - \vec{z}_2$ and $r_{12} =
|\vec{r}_{12}|$, and we used the Newtonian equation of motion. 
From this equation, we have the mass-energy relation at the
1 PN order, 
\begin{equation}
P_1^{\tau}= m_1 \left[ 1+\epsilon^2 \left( \frac12 v_1^2 
+\frac{3m_2}{r_{12}} \right) \right] +O(\epsilon^3).  
\label{1PNmass}
\end{equation}

Then we have to calculate the 1 PN order $Q_A^i$ from 
Eqs. (\ref{tLLtt6}), (\ref{tLLti6}) and (\ref{QL}).
The result is that $Q_A^i = \epsilon^2 m_A^2 v_A^i/(6 \epsilon R_A)$.
As $Q_A^i$ depends on  $R_A$,
we ignore it (see Sec. \ref{Arbitrariness}). 
As a result we obtain the momentum-velocity relation at the
1 PN order,  
 $P^{i}_A = P^{\tau}_A v_A^i + O(\epsilon^3)$ from  Eq.
(\ref{MomVelRelation}). 

Now as for $\mbox{}_4h^{ij}$, we first calculate the surface integrals
$Q_A^{ij}, R_A^{ij}, R_A^{kij}$ from Eqs. (\ref{QL}) and (\ref{RL})
using Eqs.  (\ref{tLLij4}),
(\ref{tLLtt6}), and (\ref{tLLti6}).  We then find that they depend on  
$R_A$, hence we ignore them and obtain
$$ 
h^{ij}_B = 4 \epsilon^4 \sum_{A=1,2}
\frac{m_Av_A^{i}v_A^{j}}{r_A} + O(\epsilon^5).
$$

To obtain $\mbox{}_6h^{\tau\tau}$ and $\mbox{}_4h^{ij}$,
we have to evaluate  non-compact support integrals for
$\mbox{}_6h^{\tau\tau}_{N/B}$ and $\mbox{}_4h^{ij}_{N/B}$, and
$n=2$ term in Eq. (\ref{RetardedEIREE}) for $h^{\tau\tau}$.
It can be done with the help of Laplacian inverse; $\Delta \ln r_1 =
1/r_1^2$ and $\Delta \ln S = 1/r_1r_2$ \cite{Fock}.
Using the Newtonian equation of motion we obtain
\begin{eqnarray}
h^{\tau\tau} &=& 4\epsilon^4 \sum_{A=1,2} \frac{P^{\tau}_A}{r_A}
\nonumber \\
\mbox{} &+& \epsilon^6 \left[
              -2 \sum_{A=1,2}\frac{m_A}{r_A}
			     \{(\vec{n}_A\cdot\vec{v}_A)^2-v_A^2\}
                   +2\frac{m_1 m_2}{r_{12}^2}
				   \vec{n}_{12}\cdot(\vec{n}_1-\vec{n}_2)
\right. \nonumber \\
 \mbox{} 
&+& \left.
	 7 \sum_{A=1,2}\frac{m_A^2}{r_A^2} + 14\frac{m_1 m_2}{r_1 r_2}
           -14\frac{m_1 m_2}{r_{12}} \sum_{A=1,2}\frac{1}{r_A}\right] ,
\label{htt6} \\
h^{ij} &=& 4 \epsilon^4 \sum_{A=1,2}\frac{m_A v_A^i v_A^j}{r_A}
\nonumber \\
 \mbox{} &+&
 \epsilon^4\left[ \sum_{A=1,2}\frac{m_A^2}{r_A^2}n_A^i n_A^j
-\frac{8 m_1 m_2}{r_{12}S}n_{12}^i n_{12}^j
\right.
\nonumber \\
\mbox{} &-& \left.
8\left(\delta^{i}\mbox{}_{k}\delta^{j}\mbox{}_{l}
  - \frac{1}{2}\delta^{i j}\delta_{k l}\right)
\frac{m_1 m_2}{S^2}(\vec{n}_{12}-\vec{n}_{1})^{{\scriptscriptstyle (}k}
                   (\vec{n}_{12}+\vec{n}_{2})^{l{\scriptscriptstyle )}}  
\right].
\end{eqnarray}
Here $S = r_1 + r_2 + r_{12}$, $\vec{n}_{12} = \vec{r}_{12}/r_{12}$,
$\vec{n}_{A} = \vec{r}_{A}/r_{A}$.

Inserting the field $\mbox{}_4h^{\mu\nu}$,
$\mbox{}_6h^{\tau\tau}$ into Eqs. (\ref{tLLti6}) and (\ref{tLLij6})
and evaluating the surface integrals in Eq. (\ref{generaleom}),
we obtain the 1 PN equation of motion;
\begin{eqnarray}
m_1 \frac{dv_1^i}{d\tau}=&&
 -\frac{m_1m_2}{r_{12}^2}n_{12}^i 
\nonumber\\
&&+\epsilon^2 \frac{m_1m_2}{r_{12}^2} \left[ n_{12}^i 
\left( -v_1^2-2v_2^2+\frac32 (\vec n_{12}\cdot\vec v_2)^2 
+4(\vec v_1\cdot\vec v_2)+\frac{5m_1}{r_{12}}
+\frac{4m_2}{r_{12}} \right)
\right.
\nonumber\\
&&~~~~~~~~+ \left. V^i \left( 4(\vec n_{12}\cdot\vec v_1) 
-3(\vec n_{12}\cdot\vec v_2) \right) \right] , 
\label{1PNEOMFinal}
\end{eqnarray}
where we defined the relative velocity as 
$\vec V=\vec v_1-\vec v_2$ 
and we used the Newtonian equation of motion and Eq. (\ref{1PNmass}).

Finally let us give a summary of our procedure.
With the $n-1$ PN order equation of motion and $h^{\mu\nu}$ in hand,  
we first derive the n PN evolution equation for $P^{\tau}_A$.
Then we solve it functionally and obtain the mass-energy relation
at the n PN order. Next we calculate $Q_A^i$ at the n PN order and
derive the momentum-velocity relation at the n PN order. Then
we calculate $Q_A^{Li}$ and $R_A^{Lij}$. With the n PN mass-energy
relation, the n PN momentum-velocity relation, $Q_A^{Li}$, and
$R_A^{Lij}$, we next derive the n PN  
deviation field $h^{\mu\nu}$. Finally 
we evaluate the surface integrals which appear in the right hand side 
of Eq. (\ref{generaleom}) and obtain the n PN equation of motion.
In the above calculations we use n-1 PN order equation of motion
to reduce the order, e.g., when we meet $\epsilon^2 dv^{i}_1/d\tau$ in
the right hand side of the equation motion and we have to evaluate  this 
up to $\epsilon^2$, then we use the Newtonian equation of motion and
replace it by $- \epsilon^2 m_2 r_{12}^i/r_{12}^3 $.
Basically we shall derive the 2.5 PN equation of motion
with the procedure as described above.

\section{The 1.5 PN Equation of Motion}
\label{1hPNEOM}

In the appendix \ref{RLijandQLi}, we show that 
$Q_A^{Li}$ and $R_A^{Lij}$ do not contribute to
$h^{\mu\nu}$ up to the 2.5 PN order. There, we also derive
the momentum-velocity relation up to the 1.5 PN order
$$
P_A^{i} = P_A^{\tau}v_A^i + O(\epsilon^4), 
$$
and the relations between $Z_A^{Lij}$ and some moments; the results are that
$Z_A^{ij}=\epsilon^2 P_A^{\tau}v_A^iv_A^j + O(\epsilon^4)$ and 
$Z_A^{kij} = \epsilon^2 M_A^{k(i}v_A^{j)} + O(\epsilon^4)$.
Henceforth we shall use Eqs. (\ref{hBtiInAp}) and  (\ref{hBijInAp}).  

Now let us  derive the 1.5 PN equation of motion, 
which 
is $O(\epsilon^3)$
correction to the Newtonian equation of motion. Thus
as the integrands in Eq. (\ref{EvolOfFourMom}),
we shall use $\mbox{}_7[-g t_{LL}^{\mu\nu}]$ and
hence we have to obtain  $\mbox{}_7h^{\tau\tau}$ and 
$\mbox{}_5h^{\mu i}$. The fields that we shall evaluate are  
\begin{eqnarray}
h^{\tau\tau} &=& \mbox{}_{4 \le}h^{\tau\tau} -
 4 \epsilon^7 \frac{d}{d\tau}\oint_{\pa N}dS_k
 \mbox{}_6[-g t_{LL}^{\tau k}] - 
 \frac{2}{3}
\epsilon^7 \frac{\pa^3}{\pa\tau^3}\sum_{A=1,2}P_A^{\tau}r_A^2 
+O(\epsilon^8), \\
h^{\tau i} &=& \mbox{}_{4 \le}h^{\tau i} 
-4 \epsilon^5 \oint_{\pa N} dS_k \mbox{}_4[-g t_{LL}^{i k}]
+ O(\epsilon^6),   \\
h^{ij} &=&  \mbox{}_{4 \le}h^{\tau\tau} - 
 4 \epsilon^5 \frac{d}{d\tau}\left(
					\sum_{A=1,2}
P_A^{\tau}v_A^iv_A^j + 
 			     \epsilon^4 \int_{N/B}d^3y
				 \mbox{}_4[-g t_{LL}^{ij}]\right) 
+ O(\epsilon^6). 
\end{eqnarray}
Here $\mbox{}_{\le n}h^{\mu\nu}$ denotes the field complete up to the
order $\epsilon^{n}$.

Now let us follow the procedure described in the previous section.
{}From Eqs. (\ref{tLLtt7}) and  (\ref{tLLti7}), the
mass-energy relation at the 1.5 PN order becomes  
$$
P_A^{\tau} = (P_A^{\tau})_{\le 1PN} + O(\epsilon^4)
$$
Here $(P_A^{\tau})_{\le 1PN}$ stands for the 1 PN order mass-energy
relation, Eq. (\ref{1PNmass}) (without $O(\epsilon^3)$ symbol).

A straightforward calculation for the $N/B$ contribution results in  
\begin{eqnarray}
\mbox{}_{\le 7}h^{\tau\tau} &=&
4\epsilon^4 \sum_{A=1,2} \frac{P^{\tau}_A}{r_A}
\nonumber \\
\mbox{} &+& \epsilon^6 \left[
              2 \sum_{A=1,2}\frac{\pa^2}{\pa\tau^2}
			  \left(P_A^{\tau}r_A\right)
\right. \nonumber \\
 \mbox{} 
&+& \left.
	 7 \sum_{a,b=1,2}\frac{P_a^{\tau}P_b^{\tau}}{r_a r_b}
           -14\frac{P_1^{\tau} P_2^{\tau}}{r_{12}}
		   \sum_{A=1,2}\frac{1}{r_A}\right] 
\nonumber \\ 
\mbox{} &-&
 \epsilon^7\frac{4}{3}\frac{d}{d\tau}
 \left(\sum_{A=1,2}P_A^{\tau}v_A^2
  - \frac{P_1^{\tau}P_2^{\tau}}{r_{12}}\right),
\\
\mbox{}_{\le 5}h^{\tau i} &=& 4
 \epsilon^4 \sum_{A=1,2} \frac{P_A^{\tau} v_A^i}{r_A},
\\
\mbox{}_{\le 5}h^{ij} &=&
  4 \epsilon^4 \sum_{A=1,2}\frac{P_A^{\tau} v_A^i v_A^j}{r_A}
\nonumber \\
 \mbox{} &+&
 \epsilon^4\left[ \sum_{A=1,2}\frac{P_A^{\tau 2}}{r_A^2}n_A^i n_A^j
-\frac{8 P_1^{\tau} P_2^{\tau}}{r_{12}S}n_{12}^i n_{12}^j
\right.
\nonumber \\
\mbox{} &-& \left.
8\left(\delta^{i}\mbox{}_{k}\delta^{j}\mbox{}_{l}
  - \frac{1}{2}\delta^{i j}\delta_{k l}\right)
\frac{P_1^{\tau} P_2^{\tau}}{S^2}
(\vec{n}_{12}-\vec{n}_{1})^{{\scriptscriptstyle (}k}
                   (\vec{n}_{12}+\vec{n}_{2})^{l{\scriptscriptstyle )}}  
\right]
\nonumber \\
\mbox{} &-&
2 \epsilon^5 \mbox{}^{(3)}I_{orb}^{ij}.  
\end{eqnarray}
Here 
$I_{orb}^{ij} = \sum_{A=1,2}P_A^{\tau}z_A^iz_A^j$ and
$\mbox{}^{(n)}I$ denotes $d^n I/d\tau^n$.

Note that $\mbox{}_5h^{ij}$ and $\mbox{}_7h^{\tau\tau}$ depend
only on time, thus from Eq. (\ref{tLLij7}), the equation of motion
at the 1.5 PN order becomes
$$
\frac{d v_A^i}{d\tau} = \left(\frac{d v_A^i}{d\tau}\right)_{\le 1PN}
+ O(\epsilon^4).
$$

\section{The 2 PN equation of motion}

In this and next section,
 we shall derive the 2.5 PN equation of motion using
the procedure described in the section \ref{NandFPNEOM}.
The main problem that we have to solve is to derive 
$h^{\mu\nu}_{N/B}$. Thus, from Eq. (\ref{NBcontribution}),
the problem is reduced to solve the Poisson equations 
$\Delta g(\vec x) = f(\vec{x})$. Fortunately, 
all the solutions of the Poisson equations 
that we need to derive the 2.5 PN equation of motion are
obtained in 
\cite{BFP,JS98}.       

As the first step of the 2 PN order calculation,
we have to derive $\mbox{}_6h^{\tau i}$ which will be
needed to obtain the 2 PN order evolution equation for $P_A^{\tau}$.
\begin{eqnarray}
\mbox{}_{\le 6}h^{\tau i} &=& 4
 \epsilon^4 \sum_{A=1,2} \frac{P_A^{\tau} v_A^i}{r_A}
+ 4 \epsilon^6
\int_{N/B}d^3y\frac{\mbox{}_6[-g t_{LL}^{\tau i}]}{|\vec{x}-\vec{y}|}
+ 2 \epsilon^6 \frac{\pa^2}{\pa \tau^2}\left(\sum_{A=1,2}
									   P_A^{\tau}v_A^i r_A\right)
\end{eqnarray}

Now we shall show calculations for the $N/B$ contribution in detail.
The integrand is
$$
\mbox{}_6[-16\pi g t_{LL}^{\tau i}]
= 2 \mbox{}_4h^{\tau \tau}\mbox{}_{,k}
	       \mbox{}_4h^
		   {\tau {\scriptscriptstyle [}k ,i{\scriptscriptstyle ]}}
		   - \frac{3}{4}\mbox{}_4h^{\tau k}\mbox{}_{,k}
		                \mbox{}_4h^{\tau \tau,i}.
$$
Then the first integrand contributes 
\begin{eqnarray}
\frac{2}{16 \pi}\int_{N/B}d^3y
 \frac{\mbox{}_4h^{\tau \tau}\mbox{}_{,k}
	       \mbox{}_4h^
		   {\tau {\scriptscriptstyle [}k ,i{\scriptscriptstyle ]}}}
		   {|\vec{x}- \vec{y}|} &=&
		   \frac{2}{\pi}\sum_{a,b=1,2}P_a^{\tau}P_b^{\tau}v_b^{[k}
		   \int_{N/B}d^3y\frac{r_b^{i]}r_a^k}{|\vec x - \vec
		   y|r_a^3r_b^3}
\nonumber \\
\mbox{} &=&
 2\left[\sum_{A=1,2}\frac{P_A^{\tau 2}v_A^{[k}}{r_A^2}
   \left(n_A^{i]}n_A^k - \delta^{i]k}\right)
\right. \nonumber  \\
\mbox{}   &-& \left.
   4 P_1^{\tau}P_2^{\tau}v_2^{[k}\pa_{2^{i]}}\pa_{1^{k}} {\rm ln}S
   - 4 P_1^{\tau}P_2^{\tau}v_1^{[k}\pa_{1^{i]}}\pa_{2^{k}} {\rm ln}S
  \right],  
\end{eqnarray}
where
$\pa_{A^i} = \pa/\pa z_A^i$.
In the second equality of the above equation, we used
the following formulas, 
\begin{eqnarray}
\int_{N/B}\frac{d^3y}{|\vec x - \vec y|}\frac{r_A^i r_A^j}{r_A^6} &=&
\frac{1}{8}\left(\delta^{i}\mbox{}_k\delta^{j}\mbox{}_l +
\delta^{ij}\delta_{kl}\right)\int_{N/B}d^3y
 \frac{(\Delta {\rm ln}r_A)^{,kl}}{|\vec x - \vec y|}, 
\end{eqnarray}
\begin{eqnarray}
\int_{N/B}\frac{d^3y}{|\vec x - \vec y|}
 \frac{r_1^i r_2^j}{r_1^3r_2^3} &=&
\int_{N/B}d^3y 
 \frac{1}{|\vec x - \vec y|}\frac{\pa}{\pa z_1^i}\frac{\pa}{\pa z_2^j}
(\Delta {\rm ln}S),   
\end{eqnarray}
and Eq. (\ref{NBcontribution}). 
Applying the same formulas to the other integrand, we find at last,
\begin{eqnarray}
4 \int_{N/B}d^3y
 \frac{\mbox{}_6[-g t_{LL}^{\tau i}]}{|\vec x - \vec y|}
 &=&
  \sum_{A=1,2}\frac{P_A^{\tau 2}}{r_A^2}
  \left\{(\vec{n}_A\cdot\vec v_A)n_A^i + 7 v_A^i
  \right\}
   + \frac{4 P_1^{\tau}P_2^{\tau}}{S r_{12}} (v_{1k} + v_{2k})
   \left(\delta^{ki}- n_{12}^i n_{12}^k\right)
\nonumber \\ 
\mbox{}  &&+  8 P_1^{\tau}P_2^{\tau}(v_1^i + v_2^i)
			  \left(
			   -\frac{1}{r_{12}}\sum_{A=1,2}\frac{1}{r_A}
			   + \frac{1}{r_1 r_2}
              \right)
\nonumber \\ 
\mbox{}  &&- 			   
			   16 \frac{P_1^{\tau}P_2^{\tau}}{S^2}
			   \left\{
				v_1^k(n_{12}^i - n_1^i)(n_{12}^k+n_2^k)
               +v_2^k(n_{12}^k - n_1^k)(n_{12}^i+n_2^i)
             \right\}
\nonumber \\ 
\mbox{}  &&+ 
			  12 \frac{P_1^{\tau}P_2^{\tau}}{S^2}
			   \left\{
				v_1^k(n_{12}^k - n_1^k)(n_{12}^i+n_2^i)
               +v_2^k(n_{12}^i - n_1^i)(n_{12}^k+n_2^k)
             \right\}
\end{eqnarray}			  
We show $\mbox{}_6h^{\tau i}$ in the appendix \ref{DeviationField}.

\subsection{The 2 PN evolution equation for $P^{\tau}_A$}

In this subsection we shall derive the 2 PN evolution equation
for $P_A^{\tau}$. 
The surface integrals we have to evaluate are 
\begin{eqnarray}
\frac{dP_A^{\tau}}{d\tau} &=&
 - \epsilon^2 \oint_{\partial B_A}dS_k \mbox{}_6[-g t_{LL}^{\tau k}]
 + \epsilon^2 v_A^k \oint_{\partial B_A}dS_k \mbox{}_6[-g t_{LL}^{\tau \tau}]
\nonumber \\
\mbox{} &-&
 \epsilon^4 \oint_{\pa B_A} dS_k \mbox{}_8\Lambda^{k\tau}_N
+\epsilon^{4}
v_A^k \oint_{\pa B_A} dS_k \mbox{}_8\Lambda^{\tau\tau}_N
+ O(\epsilon^5). 
\end{eqnarray}
Here $O(\epsilon^2)$ terms become   
\begin{eqnarray}
 -  \oint_{\partial B_1}dS_k \mbox{}_6[-g t_{LL}^{\tau k}]
 + v_1^k \oint_{\partial B_1}dS_k \mbox{}_6[-g t_{LL}^{\tau \tau}]
&=&
-\frac{P^{\tau}_1 P^{\tau}_2}{r_{12}^2}\left[
4(\vec{n}_{12}\cdot\vec v_1) - 3(\vec{n}_{12}\cdot\vec v_2)
									   \right].
\label{1PNfor2PNinPt}
\end{eqnarray}
This is the 1 PN order evolution equation with the
mass $m_A$ replaced by the energy $P_A^{\tau}$.

{}From the 2 PN order, $\chi^{\mu\nu\alpha\beta}_N$ comes into
play. We can see this by the definition of $P_A^{\tau}$, Eq.
(\ref{DefOfMomentum}). Noticing  
$\chi^{\tau\tau\alpha\beta}_N\mbox{}_{,\alpha\beta} =
\chi^{\tau\tau ij }_N\mbox{}_{,ij}$, we rewrite Eq.
(\ref{DefOfMomentum}) as
\begin{eqnarray}
&&
P^{\tau}_A(\tau)= \epsilon^2 \int_{B_A} d^3\alpha_A \Theta^{\tau \tau}_N  
+ \epsilon^{-4} \oint_{\pa B_A} dS_i \chi^{\tau \tau ij}_N\mbox{}_{,j}. 
\end{eqnarray}
Since $\chi^{\tau\tau\alpha\beta}_N\mbox{}_{,\alpha\beta}$ is of
order $\epsilon^8$, the surface integral of it first appears
at the order $\epsilon^4$, i.e., at the 2 PN order.

Now since $\Theta_N^{\mu\nu}$ and    
$\chi^{\mu\nu\alpha\beta}_N\mbox{}_{,\alpha\beta}$
are conserved separately, i.e.,
$\Theta_N^{\mu\nu}\mbox{}_{,\mu} = 0 = 
\chi^{\mu\nu\alpha\beta}_N\mbox{}_{,\alpha\beta\mu} $, 
 we shall deal with $\Theta_N^{\mu\nu}$ and 
$\chi^{\mu\nu\alpha\beta}_N$ separately. Thus we split $P_A^{\tau}$
into two parts; $P_{A \Theta}^{\tau}$ and $P_{A \chi}^{\tau}$.
\begin{eqnarray}
P^{\mu}_{A \Theta}(\tau) &=& \epsilon^2 \int_{B_A} d^3\alpha_A
 \Theta _{N}^{\tau \mu}, 
\\
P^{\mu}_{A\chi}(\tau) &=& \epsilon^2 \int_{B_A} d^3\alpha_A
\chi_N^{\tau\mu\alpha\beta}{}_{,\alpha\beta}. 
\end{eqnarray}
We consider the evolution equation of $P^{\mu}_{A\chi}$ in the
appendix \ref{chipart}.
The evolution equation for $P_{A \Theta}^{\tau}$ becomes 
\begin{eqnarray}
\frac{dP_{A \Theta}^{\tau}}{d\tau} &=&
 \left(\frac{d P_A^{\tau}}{d\tau}\right)_{\le 1.5 PN}
\nonumber \\
\mbox{} &-&
 \epsilon^4 \oint_{\pa B_A} dS_k \mbox{}_8[-g t_{LL}^{k\tau}]
+\epsilon^{4}
v_A^k \oint_{\pa B_A} dS_k \mbox{}_8[-g t_{LL}^{\tau\tau}].  
\end{eqnarray}
{}From Eqs. (\ref{tLLtt8}) and  (\ref{tLLti8}),
the relevant combinations of $h^{\mu\nu}$ are; 
\begin{tabbing}
$\mbox{}_4h^{\tau i,j}\mbox{}_4h^{k l,m}$, \ 
$\mbox{}_4h^{\tau \tau,i}\mbox{}_4h^{j k}\mbox{}_{,\tau}$, \  
$\mbox{}_6h^{\tau i,j}\mbox{}_4h^{\tau \tau,k}$, \
$\mbox{}_6h^{\tau\tau,i}\mbox{}_4h^{\tau j,k}$, \
$\mbox{}_4h^{\tau\tau}
\mbox{}_4h^{\tau\tau,i}\mbox{}_4h^{\tau j,k}$ \\  
$\mbox{}_4h^{\tau i} 
\mbox{}_4h^{\tau\tau,j}\mbox{}_4h^{\tau \tau,k}$
,\
$\mbox{}_6h^{\tau \tau,i}\mbox{}_4h^{\tau \tau,j}$, \
$\mbox{}_4h^{\tau i,j}\mbox{}_4h^{\tau k,l}$, \
$\mbox{}_4h^{\tau \tau,i}\mbox{}_4h^{j k,l}$, \ 
$\mbox{}_4h^{\tau\tau}
\mbox{}_4h^{\tau\tau,i}\mbox{}_4h^{\tau \tau,j}$.
\end{tabbing}
Evaluating surface integrals of these ten types of integrands,
we obtain,   
\begin{eqnarray}
 -  \oint_{\partial B_1}dS_k \mbox{}_8[-g t_{LL}^{\tau k}]
&=&
-\frac{m_1 m_2}{r_{12}^2}
\left[
                \frac{9 (\vec{n}_{12}\cdot\vec v_2)^3}{2}
               -\frac{5 (\vec{n}_{12}\cdot\vec v_1)
			   (\vec{n}_{12}\cdot\vec{v}_{2})^2}{2}
               + \frac{4 v_1^2 (\vec{n}_{12}\cdot\vec{v}_{1})}{15}
\right. \nonumber \\
\mbox{} &+& \left. 
                3 (\vec{v}_1\cdot\vec{v}_2)
			   (\vec{n}_{12}\cdot\vec{v}_{2})
               - \frac{4 (\vec{v}_1\cdot\vec{v}_2)
			   (\vec{n}_{12}\cdot\vec{v}_{1})}{3}
\right. \nonumber \\
\mbox{} &-& \left.
               \frac{7 v_2^2 (\vec{n}_{12}\cdot\vec{v}_{2})}{2}
			   +\frac{7 v_2^2 (\vec{n}_{12}\cdot\vec{v}_{1})}{6}
			\right]
\nonumber \\
\mbox{} &-& \frac{m_1 m_2^2}{r_{12}^3}
 \left[-2 (\vec{n}_{12}\cdot\vec{v}_2)
			      -\frac{2 (\vec{n}_{12}\cdot\vec{v}_1)}{3}\right]
\nonumber \\
\mbox{} &-&
 \frac{m_1^2 m_2}{r_{12}^3}\left[
							\frac{19 (\vec{n}_{12}\cdot\vec{v}_2)}{2}
                           -\frac{31 (\vec{n}_{12}\cdot\vec{v}_1)}{6}  
                           -\frac{7 (\vec{n}_{12}\cdot\vec{V})}{2}
						   \right], 
\end{eqnarray}
and 
\begin{eqnarray}
v_1^k \oint_{\partial B_1}dS_k \mbox{}_8[-g t_{LL}^{\tau \tau}]
&=&
\frac{m_1 m_2}{r_{12}^2}\left[
                         -v_1^2 (\vec{n}_{12}\cdot\vec{v}_2)
						 +\frac{7 (\vec{n}_{12}\cdot\vec{v}_1)
						  (\vec{n}_{12}\cdot\vec{v}_2)^2}{2}
						 +\frac{4 v_1^2
						 (\vec{n}_{12}\cdot\vec{v}_1)}{15}
\right. \nonumber \\
\mbox{}  &-& \left.  (\vec{v}_1\cdot\vec{v}_{2})
						  (\vec{n}_{12}\cdot\vec{v}_2)
						 +\frac{8  (\vec{v}_1\cdot\vec{v}_{2})
						 (\vec{n}_{12}\cdot\vec{v}_1)}{3}
						 -\frac{5 v_2^2
						 (\vec{n}_{12}\cdot\vec{v}_1)}{6}
						\right]
\nonumber \\
\mbox{} &+& \frac{4 m_1 m_2^2 (\vec{n}_{12}\cdot\vec{v}_1)}{3}
\nonumber \\
\mbox{}  &+& \frac{28 m_1^2 m_2 (\vec{n}_{12}\cdot\vec{v}_1)}{3}. 
\end{eqnarray}
Combing these results, the evolution equation for $P_A^{\tau}$ becomes
\begin{eqnarray}
\left(\frac{d P_{1 \Theta}^{\tau}}{d \tau}\right)_{\le 2PN} &=&
- \epsilon^2 \frac{m_1 m_2}{r_{12}^2}\left[
4(\vec{n}_{12}\cdot\vec{v}_1) - 3(\vec{n}_{12}\cdot\vec{v}_2)
									   \right]
\nonumber \\
\mbox{} &+&
 \epsilon^4 \frac{m_1 m_2}{r_{12}^2}\left[
              -\frac{9 (\vec{n}_{12}\cdot\vec{v}_2)^3}{2}
			  + \frac{v_1^2 (\vec{n}_{12}\cdot\vec{v}_2)}{2}
			  + 6 (\vec{n}_{12}\cdot\vec{v}_1)
			  (\vec{n}_{12}\cdot\vec{v}_2)^2
              - 2 v_1^2 (\vec{n}_{12}\cdot\vec{v}_1)
\right. \nonumber \\
\mbox{} &+& \left.
			   4 (\vec{v}_1\cdot\vec{v}_{2})
			  (\vec{n}_{12}\cdot\vec{V})
			  + 5 v_2^2 (\vec{n}_{12}\cdot\vec{v}_2)
			  -4 v_2^2 (\vec{n}_{12}\cdot\vec{v}_1)
 									\right]
\nonumber \\
\mbox{} &+&
 \epsilon^4 \frac{m_1 m_2^2}{r_{12}^3}
 [-10 (\vec{n}_{12}\cdot\vec{v}_1) + 11 (\vec{n}_{12}\cdot\vec{v}_2)]
\nonumber \\
\mbox{} &+&
 \epsilon^4 \frac{m_1^2 m_2}{r_{12}^3}
 [-4 (\vec{n}_{12}\cdot\vec{v}_2)
 + 6 (\vec{n}_{12}\cdot\vec{v}_1)].
\label{SPNEMSurfaceIntegralFormTimeResult}
\end{eqnarray}

We can functionally integrate this equation and the result is
\begin{eqnarray}
P^{\tau}_{1 \Theta} &=& m_1 \left( 1 + \epsilon^2 \mbox{}_2\Gamma_1
                            + \epsilon^4 \mbox{}_4\Gamma_1
				   \right) + O(\epsilon^5).
\label{2PNEnergy}
\end{eqnarray}
Here 
\begin{eqnarray}
\mbox{}_2\Gamma_1 &=& \frac{1}{2}v_1^2 + \frac{3 m_2}{r_{12}},
\label{DefOfGamma2}\\
\mbox{}_4\Gamma_1 &=&
- \frac{3 m_2}{2 r_{12}} (\vec{n}_{12}\cdot\vec{v}_2)^2
+ \frac{2 m_2}{r_{12}}v_2^2
+ \frac{7 m_2}{2 r_{12}}v_1^2
- \frac{4 m_2}{r_{12}} (\vec{v}_1\cdot\vec{v}_{2})
+ \frac{3}{8} v_1^4
+ \frac{7 m_2^2}{2 r_{12}^2}
- \frac{5 m_1 m_2}{2 r_{12}^2}.
\label{DefOfGamma4}
\end{eqnarray}  
Eqs. (\ref{Ptchi2hPN}) and (\ref{2PNEnergy}) give 
the 2 PN order mass-energy relation.

\subsection{The 2 PN gravitational field}

To derive the 2 PN equation of motion, we need 
$\mbox{}_8h^{\tau\tau} + \mbox{}_6h^{l}\mbox{}_l$ (see Eqs.
(\ref{tLLij8}) and (\ref{generaleom})).   

The field $h^{\mu\nu}$ that we shall evaluate from now on are as follows.
\begin{eqnarray}
h^{\tau\tau} &=& \mbox{}_{\le 7}h^{\tau\tau} +
 4 \epsilon^8 \int_{N/B}d^3y
 \frac{\mbox{}_8\Lambda_N^{\tau\tau}}{|\vec x - \vec y|}
 + 2\epsilon^8 \frac{\pa^2}{\pa \tau^2}
  \left(\int_{N/B}d^3y |\vec x - \vec y|
   \mbox{}_6[-g t_{LL}^{\tau\tau}]\right)
\nonumber \\
 \mbox{} &+&
   \frac{1}{6}\epsilon^8
  \frac{\pa^4}{\pa \tau^4}\left(\sum_{A=1,2}P_A^{\tau}r_A^3\right)
  + O(\epsilon^9),
\\
h^{ij} &=& \mbox{}_{\le 5}h^{ij} + 4 \epsilon^6
 \int_{N/B} 
 \frac{\mbox{}_6[-g t_{LL}^{ij}]}{|\vec x - \vec y|}
 \nonumber \\
 \mbox{} &+& 
 2 \epsilon^6 \frac{\pa^2}{\pa \tau^2}
 \left(\int_{N/B}d^3y |\vec x - \vec y|\mbox{}_4[(-g)t_{LL}^{ij}]
  + \sum_{A=1,2}
  P_A^{\tau}v_A^iv_A^j r_A \right) +O(\epsilon^7).
\end{eqnarray}
Then $h^{\tau\tau} + \epsilon^2 h^{l}\mbox{}_l -
(\mbox{}_{\le 7}h^{\tau\tau}
+ \epsilon^2 \mbox{}_{\le 5}h^{l}\mbox{}_l)$ gives
\begin{eqnarray}
\mbox{} &&
 4 \int_{N/B}d^3y
\frac{\mbox{}_8\Lambda_N^{\tau\tau}
+ \mbox{}_6[-g t_{LL}\mbox{}^k\mbox{}_k]}
{|\vec x - \vec y|}
\nonumber \\
\mbox{} &+& 2 \frac{\pa^2}{\pa \tau^2}\left[\sum_{A=1,2}P_A^{\tau}v^2_A r_A 
			+ \int_{N/B}d^3y |\vec x - \vec y|
			\left(\mbox{}_6[-g t_{LL}^{\tau\tau}] +
			\mbox{}_4[-g t_{LL}\mbox{}^k\mbox{}_k]\right)
									  \right]
\nonumber \\
\mbox{} &+& \frac{1}{6}\frac{\pa^4}{\pa \tau^4}
 \sum_{A=1,2}\left(P_A^{\tau}r_A^3\right).
\end{eqnarray}
The integrand 
$\mbox{}_8\Lambda_N^{\tau\tau} + \mbox{}_6[-g t_{LL}\mbox{}^k\mbox{}_k]$
contains 
\begin{eqnarray}
 16 \pi \mbox{}_8\Lambda_N^{\tau\tau} +
 \mbox{}_6[-g t_{LL}\mbox{}^k\mbox{}_k]
 &=& 
 2 \mbox{}_4h^{\tau i}\mbox{}_4h^{\tau j}\mbox{}_{,ij}
 - \mbox{}_4h^{ij}\mbox{}_4h^{\tau\tau}\mbox{}_{,ij}
 - \mbox{}_4h^{\tau\tau}\mbox{}_4h^{ij}\mbox{}_{,ij}
 \nonumber \\
 \mbox{} &+& 2 \mbox{}_4h^{\tau k}\mbox{}_{,l}
  \mbox{}_4h^{\tau l}\mbox{}_{,k}
  - \frac{1}{2}\mbox{}_4h^{\tau k}\mbox{}_{,k}
  \mbox{}_4h^{\tau l}\mbox{}_{,l}
\nonumber \\
\mbox{} &-& 2\mbox{}_4h^{\tau\tau,k}\mbox{}_6h^{\tau\tau}\mbox{}_{,k}
 + \frac{9}{8}\mbox{}_4h^{\tau\tau}\mbox{}_4h^{\tau\tau,k}
 \mbox{}_4h^{\tau\tau}\mbox{}_{,k}.
\label{Lambdatt8PlustLLkk6}
\end{eqnarray}
The evaluation of the first three terms may call for great care.
Let us show the results.
\begin{eqnarray}
\int_{N/B}\frac{- d^3y}{4 \pi |\vec x - \vec y|}
(-2 \mbox{}_4h^{\tau i}\mbox{}_4h^{\tau j}\mbox{}_{,ij})
&=&
8\sum_{A=1,2}  \frac{P_A^{i}P_A^j}{r_A^2}
\left(3 n_A^in_A^j - \delta^{ij}\right)
\nonumber \\
\mbox{} &-& 32 P_1^{i}P_2^j
\left(\pa_{1^i}\pa_{1^j}+ \pa_{2^i}\pa_{2^j} \right)\ln S
\nonumber \\
\mbox{} 
&+& \frac{32(\vec P_1\cdot \vec P_2)}{3 r_{12}}\sum_{A=1,2}\frac{1}{r_{A}},
\end{eqnarray}
where the last term comes from the boundary term
(surface integral term)  
in Eq.
(\ref{NBcontribution}).
\begin{eqnarray}
\int_{N/B}\frac{-d^3y}{4 \pi |\vec x - \vec y|}
\mbox{}_4h^{ij}\mbox{}_4h^{\tau\tau}\mbox{}_{,ij}
&=&
4 \frac{P_1^{\tau 2}v_1^iv_1^j}{r_1^2}
\left(\delta^{ij}- 3 n_1^i n_1^j\right)
\nonumber \\
\mbox{}  
&+& 16 P_1^{\tau}P_2^{\tau}v_1^iv_1^j \pa_{2^i}\pa_{2^j}\ln S
\nonumber \\
\mbox{} &+& \frac{4}{3}\frac{P_1^{\tau 3}}{r_1^3}
+ P_1^{\tau 2}P_2^{\tau}
\left(-16 H_1 - K_1\right)
\nonumber \\
\mbox{} 
&-&
 \frac{16 P_1^{\tau}P_2^{\tau}v_1^2}{3}
+ 4 P_1^{\tau 2} P_2^{\tau}
\left(-\frac{4}{3 r_1 r_{12}^2} + \frac{1}{6 r_2 r_{12}^2}\right)
+(1 \leftrightarrow 2), 
\label{ExampleOf334}
\end{eqnarray}
where $(1 \leftrightarrow 2)$ means that there are the same terms
but with the labels 1 and 2 exchanged. 
The last two terms arise from the body zone boundary term in the
Eq. (\ref{NBcontribution}).  
$H_1$ and $K_1$ have been derived by Blanchet, Damour, and
Iyer\cite{BDI95}.
They satisfy the following poisson equations in the sense of usual
function in N/B;
\begin{eqnarray}
&&
\Delta K_{1}  = 2\partial_{i}\partial_{j}\left(\frac{1}{r_{2}}\right)
\partial_{i}\partial_{j}(\ln r_{1}),  \\ 
&&
\Delta H_{1} =  2\partial_{i}\partial_{j}\left(\frac{1}{r_{1}}\right)
\pa_{1^i}\pa_{2^j}\ln S.
\end{eqnarray}
Their explicit expressions are 
\begin{eqnarray}
&&
K_{1}
= -\frac{1}{r_2^3} + \frac{1}{r_2 r_{12}^2} - \frac{1}{r_{1}^2 r_2}
+ \frac{r_2}{2 r_{1}^2 r_{12}^2} + \frac{r_{12}^2}{2 r_{1}^2 r_{2}^3}
+ \frac{r_{1}^2}{2 r_{2}^3 r_{12}^2}, \\
&&
H_{1}
= -\frac{1}{2 r_1^3} - \frac{1}{4 r_{12}^3} - \frac{1}{4 r_{1}^2 r_{12}}
- \frac{r_2}{2 r_{1}^2 r_{12}^2} + \frac{r_{2}}{2 r_{1}^3 r_{12}}
+ \frac{3 r_{2}^2}{4 r_{1}^2 r_{12}^3}  +
\frac{r_{2}^2}{2 r_{1}^3 r_{12}^2}
- \frac{r_{2}^3}{2 r_{1}^3 r_{12}^3}.   
\end{eqnarray}
We note that they go to zero when $r \rightarrow \infty$.
The third term of Eq. (\ref{Lambdatt8PlustLLkk6})
can be evaluated as 
\begin{eqnarray}
\int_{N/B}\frac{-d^3 y}{4 \pi |\vec x - \vec y|} 
 \mbox{}_4h^{\tau\tau}\mbox{}_4h^{ij}\mbox{}_{,ij}
&=&
2 \sum_{A=1,2}P_A^{\tau 2}v_A^iv_A^j
(3\pa_{A^i}\pa_{A^j} - \delta^{ij} \Delta_{A})\ln r_A
\nonumber \\
\mbox{} &+& 8 \sum_{A=1,2} P_A^{\tau 2} a_A^k\pa_{A^k}\ln r_A
\nonumber \\
\mbox{} &+& 16 P_1^{\tau}P_2^{\tau}
 (v_1^kv_1^l \pa_{1^k}\pa_{1^l} +
 v_2^kv_2^l \pa_{2^k}\pa_{2^l})\ln S
\nonumber \\
\mbox{} &+& 16 P_1^{\tau}P_2^{\tau}
 (a_1^k \pa_{1^k} + a_2^k \pa_{2^k}) \ln S
\nonumber \\
\mbox{} &-& \frac{16 P_1^{\tau}P_2^{\tau}}{3 r_{12}}\sum_{A=1,2}
 \frac{v_A^2}{r_A},  
\end{eqnarray}
 of the star A,
$$
a^i_A = \frac{d v_A^i}{d\tau}
$$
The last term again comes from the boundary term in Eq.
(\ref{NBcontribution}). 

The complete $h^{\mu\nu}$ 
are shown in the appendix \ref{DeviationField}.

\subsection{The 2 PN equation of motion}

To derive the 2 PN equation of motion, we have to
derive the momentum-velocity relation valid up to
$O(\epsilon^4)$.
Then, when the integrand is $\mbox{}_8\Lambda_N^{\tau\mu}$, 
$Q_A^{i}$ becomes
\begin{eqnarray}
Q_A^i &=& Q^i_{A LL} + Q^i_{A \chi}
\\
Q_{1 LL}^{i} &=& \frac{m_1^2 m_2 v_1^i}{3 \epsilon R_1 r_{12}} -
 \frac{m_1^2 v_1^2 v_1^i}{15 \epsilon R_1} -
 \frac{2 m_1^2 m_2 v_2^i}{3 \epsilon R_1 r_{12}},
\\
 Q_{A \chi}^{i} &=& 0
\end{eqnarray}
where $Q_{A LL}^i$ and $Q_{A \chi}^i$ come from the
integrands 
$-g t_{LL}^{\mu\nu}$ and $\chi^{\mu\nu\alpha\beta}_N\mbox{}_{,\alpha\beta}$.
Thus the momentum-velocity relation becomes
\begin{eqnarray}
P_{A \Theta}^{i} &=& P_{A \Theta}^{\tau}v_A^i + O(\epsilon^5), \\
P_{A \chi}^{i} &=& P_{A \chi}^{\tau}v_A^i + O(\epsilon^5).
\end{eqnarray}

Now, by the divergencefree nature of 
$\chi^{\mu\nu\alpha\beta}_N\mbox{}_{,\alpha\beta}$,
$P_{A \chi}^{\mu}$ is itself conserved (see the appendix \ref{chipart}).
Hence we shall consider
$P_{A \Theta}^{\mu}$ only.
Up to the 2 PN order, we have to evaluate 
\begin{eqnarray}
\frac{d P_{A \Theta}^i}{d\tau} &=& - \sum_{n=0}^{2}
 \epsilon^{2n}
 \oint_{\partial B_A}dS_k \mbox{}_{4+2n}[-g t_{LL}^{i k}]
 + \sum_{n=1}^{2}
 \epsilon^{2n} v_A^k
 \oint_{\partial B_A}dS_k \mbox{}_{4+2n}[-g t_{LL}^{i\tau}]
+ O(\epsilon^5). 
\end{eqnarray}
Evaluating the above surface integrals and using the mass-energy
relation valid up to the 2 PN order,
we obtain the 2 PN equation of motion.
We show it and the 2.5 PN order correction in the next section.

\section{The 2.5 PN equation of motion}

In this section we shall derive the 2.5 PN equation of motion.

Now we first determine the mass-energy relation up to the 2.5 PN order.
The surface integrals we have to evaluate are 
\begin{eqnarray}
\frac{d P_{A \Theta}^{\tau}}{d\tau} &=&
 \left(\frac{d P_{A \Theta}^{\tau}}{d\tau}\right)_{\le 2 PN}
\nonumber \\
\mbox{} &-&
 \epsilon^5 \oint_{\pa B_A} dS_k \mbox{}_9[-g t_{LL}^{k\tau}]
+\epsilon^{5}
v_A^k \oint_{\pa B_A} dS_k  \mbox{}_9[-g t_{LL}^{\tau\tau}]. 
\end{eqnarray}
The evolution equation for $P_{A \chi}^{\tau}$ is shown
in the appendix \ref{chipart}.
Evaluating these surface integrals, we find that 
$$
\left(\frac{d P_{A \Theta}^{\tau}}{d\tau}\right)_{2.5 PN} = 0.
$$ 
Thus we have the mass-energy
relation  (of $P_{A \Theta}^{\tau}$ part)
\begin{eqnarray}
P_{1 \Theta}^{\tau} &=&
 m_1\left[1 + \epsilon^2 \mbox{}_2\Gamma_1 + \epsilon^4
	 \mbox{}_4\Gamma_1\right]  + O(\epsilon^6).
\label{2hPNMassEnergy}
\end{eqnarray} 
Here the definition of $\mbox{}_n\Gamma_A$ are given by Eqs.
(\ref{DefOfGamma2}) and (\ref{DefOfGamma4}).  We discuss 
an interesting interpretation of this mass-energy relation in
the appendix \ref{MeaningOfPt}.

Next we have to derive the momentum-velocity relation. By evaluating
explicitly, we find that $Q_A^i = 0$. Hence we have
$$ 
P_A^{i} = P_A^{\tau}v_A^i + O(\epsilon^{6}).
$$

Finally we have to derive the deviation field $h^{\mu\nu}$ up to
the 2.5 PN order; $\mbox{}_9h^{\tau\tau}$, $\mbox{}_7h^{\mu\nu}$. 
We have
\begin{eqnarray}
\epsilon^7\mbox{}_7h^{\tau\tau} +
 \epsilon^9\mbox{}_9h^{\tau\tau} &=&
 - \frac{4}{3}\epsilon^7\frac{d}{d \tau}
 \left[\sum_{A=1,2}P_A^{\tau}v_A^2 - \frac{P_1^{\tau}P_2^{\tau}}{r_{12}}
 \right]
\nonumber \\
\mbox{} &+& 4\epsilon^9 \mbox{}^{(3)}I_{orb}^{ij}
 \sum_{A=1,2}
 \frac{m_A}{r_A}
 \left[n_A^in_A^j-\frac{1}{3}\delta^{ij}\right]
\nonumber \\
\mbox{} &-& \frac{1}{30}\epsilon^9\frac{\pa^5}{\pa \tau^5}
 \left[\sum_{A=1,2}m_A r_A^4\right]
 \nonumber \\  
\mbox{} &+& ({\rm irrelevant \,\,\,\,terms}),   
 \\
\epsilon^7 \mbox{}_7h^{\tau i} &=& - \epsilon^7 \frac{2}{3}
 \frac{\pa^3}{\pa \tau^3}\sum_{A=1,2} m_Av_A^ir_A^2, 
 \\
\epsilon^5 \mbox{}_5h^{l}\mbox{}_l +
\epsilon^7 \mbox{}_7h^{l}\mbox{}_l  &=&
-2 \epsilon^5 \mbox{}^{(3)}I^l_{orb}\mbox{}_l
\nonumber \\
\mbox{} &-& \epsilon^7\frac{2}{3}\frac{\pa^3}{\pa \tau^3}
 \left[\sum_{A=1,2} m_A v_A^2 r_A^2 -
  \frac{m_1 m_2}{2 r_{12}}\sum_{A=1,2}r_A^2\right]
\nonumber \\
 \mbox{} &+&  ({\rm irrelevant \,\,\,\,terms}).
\end{eqnarray}

Evaluating the surface integrals in Eq. (\ref{generaleom})
with  the deviation field above,
we at last obtain the 2.5 PN order equation motion.  
\begin{eqnarray}
m_1 \frac{dv_1^i}{d\tau} &=&
 - \frac{m_1m_2}{r_{12}^2}n_{12}^i 
\nonumber\\
&&
+ \epsilon^2 \frac{m_1m_2}{r_{12}^2} n_{12}^i 
\left[ -v_1^2-2v_2^2 +4(\vec v_1\cdot\vec v_2)
+\frac32 (\vec n_{12}\cdot\vec v_2)^2 
+\frac{5m_1}{r_{12}}
+\frac{4m_2}{r_{12}} \right]
\nonumber\\
&&
+ \epsilon^2 \frac{m_1m_2}{r_{12}^2}
V^i \left[ 4(\vec n_{12}\cdot\vec v_1) 
-3(\vec n_{12}\cdot\vec v_2)  \right]
\nonumber \\
&&
+\epsilon^4 \frac{m_1 m_2}{r_{12}^2} n_{12}^i
\left[
-2 v_2^4 + 4 v_2^2 (\vec{v}_1 \cdot \vec{v}_2 )
       - 2 (\vec{v}_1 \cdot \vec{v}_2)^2
       + \frac32 v_1^2 (\vec{n}_{12}\cdot\vec{v}_2)^2
       + \frac92 v_2^2 (\vec{n}_{12}\cdot\vec{v}_2)^2
\right.
\nonumber \\
&&~~~~~~~~~~~~~~~~~~~
       - \left. 6 (\vec{v}_1 \cdot \vec{v}_{2})
           (\vec{n}_{12}\cdot\vec{v}_2)^2
       - \frac{15}{8}(\vec{n}_{12}\cdot\vec{v}_2)^4
\right.
\nonumber \\
&&~~~~~~~~~~~~~~~~~~~
+ \left. \frac{m_1}{r_{12}}
\left(
       -\frac{15}{4}v_1^2 + \frac54 v_2^2
       - \frac52 (\vec{v}_1 \cdot \vec{v}_{2})
\right. \right. 
\nonumber \\
&&~~~~~~~~~~~~~~~~~~~~~~~~~~~~~
       +\left. \left. \frac{39}{2} (\vec{n}_{12}\cdot\vec{v}_1)^2
       - 39 (\vec{n}_{12}\cdot\vec{v}_1)(\vec{n}_{12} \cdot \vec{v}_{2})
       + \frac{17}{2} (\vec{n}_{12}\cdot\vec{v}_2)^2
\right)
\right. 
\nonumber \\
&&~~~~~~~~~~~~~~~~~~~
+ \left. \frac{m_2}{r_{12}}
\left(
       4v_2^2  - 8 (\vec{v}_1 \cdot \vec{v}_{2})
\right. \right. 
\nonumber \\
&&~~~~~~~~~~~~~~~~~~~~~~~~~~~~~
       + \left. \left. 2 (\vec{n}_{12}\cdot\vec{v}_1)^2
       - 4 (\vec{n}_{12}\cdot\vec{v}_1)(\vec{n}_{12} \cdot \vec{v}_{2})
       + 6 (\vec{n}_{12}\cdot\vec{v}_2)^2
\right)
\right.
\nonumber \\
&&~~~~~~~~~~~~~~~~~~~
- \left. \frac{57}{4}\frac{m_1^2}{r_{12}^2}
   - 9 \frac{m_2^2}{r_{12}^2}
   - \frac{69}{2} \frac{m_1 m_2}{r_{12}^2}
\right]
\nonumber \\
&&
+\epsilon^4 \frac{m_1 m_2}{r_{12}^2} V^i
\left[
v_1^2 (\vec{n}_{12}\cdot\vec{v}_2)
+ 4 v_2^2 (\vec{n}_{12}\cdot\vec{v}_1)
- 5 v_2^2 (\vec{n}_{12}\cdot\vec{v}_2)
- 4 (\vec{v}_1 \cdot \vec{v}_{2})(\vec{n}_{12}\cdot\vec{V})
\right.
\nonumber \\
&&~~~~~~~~~~~~~~~~~~~
- \left.6 (\vec{n}_{12}\cdot\vec{v}_1)(\vec{n}_{12}\cdot\vec{v}_2)^2
+ \frac92 (\vec{n}_{12}\cdot\vec{v}_2)^3
\right.
\nonumber \\
&&~~~~~~~~~~~~~~~~~~~
+ \left.\frac{m_1}{r_{12}}
\left(
- \frac{63}{4}(\vec{n}_{12}\cdot\vec{v}_1)
+ \frac{55}{4}(\vec{n}_{12}\cdot\vec{v}_2)
\right)
\right.
\nonumber \\
&&~~~~~~~~~~~~~~~~~~~
+ \left.\frac{m_2}{r_{12}}
\left(
- 2(\vec{n}_{12}\cdot\vec{v}_1)
- 2(\vec{n}_{12}\cdot\vec{v}_2)
\right)
\right]
\nonumber \\
&&
+\epsilon^5 \frac{4 m_1^2 m_2}{5 r_{12}^3} 
\left[n_{12}^i
(\vec{n}_{12}\cdot\vec{V})
\left(
-6\frac{m_1}{r_{12}} + \frac{52}{3}\frac{m_2}{r_{12}} + 3 V^2
\right)
+ V^i
\left(
2 \frac{m_1}{r_{12}} - 8 \frac{m_2}{r_{12}} - V^2
\right)
\right].
\label{2.5PNEOMFinal}
\end{eqnarray}
This is the 2.5 PN equation of motion for the relativistic compact
binary. 
We used the harmonic coordinate to derive this equation.  
This equation perfectly agrees with
the previous works\cite{DD81a,DD81b,Kopejkin,BFP}. 
Thus it is found that up to the 2.5 PN order the post-Newtonian 
equation of motion is applicable to the binary system even 
whose constituent stars have strong internal gravity.

\section{conclusion and discussion}

In this paper, we derived the equation of motion for two
compact bodies accurate through the 2.5 post-Newtonian order,
where the radiation reaction effect first appears. 
In the paper I, we have derived 
the spin-orbit and quadrupole-orbit coupling forces
which are the only multipole-orbit coupling forces up to the 2.5
PN order in our ordering. Combing these multipole-orbit coupling
forces and the result in this paper, we obtain an equation of motion 
appropriate to an inspiralling binary star system 
up to the 2.5 PN order.  
We have derived all these accelerations in the harmonic coordinate.

To deal with a strongly self-gravitating object
such as  a neutron star, 
we used the general form of the equation of motion with 
the strong filed point particle limit. 
We imposed the scalings on matter and field variables on the initial
hypersurface. This scalings enabled us to introduce a self-gravitating 
point-particle and ensured the (quasi-)stationarity of the stars. And 
in turn, the (quasi-)stationarity ensured that the internal motion 
of the star is stimulated primary by the tidal effect.
In our method, 
we did not {\it a priori} assume a geodesic equation, but
used only  the local conservation law of the total stress
energy-momentum tensor (matter's stress-energy tensor plus
Landau-Lifshitz stress-energy pseudo-tensor). Using this
conservation law, we introduced the general form of the
equation of motion which is expressed entirely in
surface integrals. The evaluation of
these integrals were made explicitly. 
The resulting 2.5 PN equation of motion agrees with 
Damour and Deruelle 2.5 PN equation of motion
\cite{DD81a,DD81b,Kopejkin,BFP}.  
Hence we find that the 2.5 PN equation of motion
is applicable to a relativistic compact binary, 
modulo the scalings 
imposed initially on the matter and field variables.

Throughout our derivation, the evolution equation for
the time component of the four momentum of the star A, $P_A^{\tau}$,
has played an important role.
As a by-product, we obtain an interesting and natural relation
between $P_A^{\tau}$ and
the binary's characteristics, Eq. (\ref{MassEnergyRelationFull}).

We ignored the dependence on the body zone size $R_A$ in the 
estimation of  the field,
the multipole moments of the compact bodies, and
the equation of motion. 
Since 
we introduced $R_A$ by hand, 
it is reasonable to
expect that the field does not depend on $R_A$.
We consider multipole moments
as renormalized moments.
It is proved that whatever $R_A$ we take as the radius of
the body zone,  
the general equation of motion does not depend on $R_A$.    

Discarding $R_A$ dependent terms when calculating the field is
similar
to regularizations such as Hadamard partie finie. Then, 
 there is a possibility that at the 3 PN order our method
might give 
$\log (r_A/\epsilon R_A)$ type ambiguity that might correspond to
the one reported in \cite{JS99,BFY2Ka,BFY2Kb}. If such log term should 
arise we could not discard $\epsilon R_A$ even practically. 

Our method, however, still differs from other works \cite{JS99,BFY2Ka,BFY2Kb}
in some points.
Thus, besides as a check 
for the 3 PN equation of motion
 obtained in \cite{JS99,BFY2Ka,BFY2Kb}, it is interesting 
to derive it to study the log type ambiguity.
 We will be  tackling the  problem and
try to derive the 3 PN equation of motion.

\section*{Acknowledgments}

Y.I. would like to acknowledge the support from
Japan Society for the Promotion of Science. This work
was supported in part by a Japanese Grant-in-Aid for
Scientific Research from the Ministry of Education,
Science and Culture, No. 11740130 (H. A.).

\appendix

\section{$Q_A^{Li}$ and $R_A^{Lij}$}
\label{RLijandQLi}

In $h_{B n=0}^{\mu\nu}$, Eqs. (\ref{hBti}), and
(\ref{hBij}),
 the moments $J_A^{Li}$ and $Z_A^{Lij}$ appear formally at the order
$\epsilon^{2 l + 4}$ and $\epsilon^{2 l + 2}$. Thus $Q^{Li}_A$ and 
$R_A^{Lij}$ appear as  
$$
h^{\tau i}_{B n=0} \sim
\cdot\cdot\cdot + 
\epsilon^4 \frac{r_A^L \hat Q^{Li}}{r_{A}^{2 l + 1}} + \cdot\cdot\cdot,
$$
$$
h^{ij}_{B n=0} \sim
\cdot\cdot\cdot + 
\epsilon^4 \frac{r_A^L \hat R^{Lij}}{r_{A}^{2 l + 1}} + 
\cdot\cdot\cdot, 
$$
where we omitted irrelevant terms and numerical coefficients.
These equations seem to suggest that $Q_A^{Li}$ and $R_A^{Lij}$
for any $l$ contribute to $h^{\mu\nu}$ at the order $\epsilon^4$. 

We solve this problem by declaring that $h^{\mu\nu}$ does not
depend on the size of the body zone boundary, $\epsilon R_A$
(see Sec. \ref{Arbitrariness}). By discarding the
terms depending on $\epsilon R_A$, we find that a finite number
of $Q_A^{Li}$ and $R_A^{Lij}$ come into play. 

Now, let us consider the n PN order  integrand
$\epsilon^{2n+2}\mbox{}_{2n+2}\Lambda_N^{\mu\nu}$ (where the n PN
order integrand means the integrand which appears at the n PN order
in the evaluation of $h^{\mu i}_{B n=0}$. For example, 
$\mbox{}_4\Lambda^{ij}_N$ is 1 PN integrand and
$\mbox{}_6\Lambda^{\mu i}$
is 2 PN integrand). 
These integrands, say,  for the star A, schematically have forms 
such as 
\begin{eqnarray}
\epsilon^{2n+2}\mbox{}_{2n + 2}\Lambda_N^{\tau i}  &\sim&
 \epsilon^{2n+2}\left(\frac{m}{r}\right)^s \frac{m^2}{r^4} 
\lambda_A^{\tau i}(\vec{n}_A\cdot\vec{n}_{AC}), 
\label{Int1InAppA}
\end{eqnarray}
\begin{eqnarray}
\epsilon^{2n+2}\mbox{}_{2n + 2}\Lambda_N^{ij}  &\sim&
 \epsilon^{2n+2}\left(\frac{m}{r}\right)^{t} \frac{m^2}{r^4}
\lambda_A^{ij}(\vec{n}_A\cdot\vec{n}_{AC}), 
\label{Int2InAppA}
\end{eqnarray}
where the label C denotes the companion star (if $A=1$, $C=2$). 
In these integrands, we expanded $r_C$ around the center of the star A, 
and  expressed the angular dependence of the integrands by the function 
$\lambda_A^{\mu\nu}$ whose forms are irrelevant here.  
$s = 0,1,\cdot\cdot\cdot [n-2]$ ($s=0$ if $n=0,1$) and
$t= 0,1,\cdot\cdot\cdot [n-1]$ ($t=0$ if $n=0$),
(where $[a]$ ($a$: real number) 
denotes the biggest integer smaller than $a$). 
$m^k$ and $r^k$ are $m_1^{l}m_2^{n}$ and $r_1^l r_{12}^n$ respectively  
where $k$, $l$, and $n$ are positive integers and $l+n = k$.  
Then from the definition  of $Q_A^{Li}$ and $R_A^{Lij}$, Eqs.
(\ref{QL}) and (\ref{RL}), Eqs. (\ref{Int1InAppA}) and 
(\ref{Int2InAppA}) give 
$$
\epsilon^4 Q_A^{Li} \sim (\epsilon R_A)^{l-s-1}\epsilon^{2n+2}  , 
$$
$$
\epsilon^4 R_A^{Lij} \sim (\epsilon R_A)^{l-t-1}\epsilon^{2n+2},  
$$
where only the $\epsilon$ and $\epsilon R_A$ dependence are shown.
Now from the argument in Sec. \ref{Arbitrariness}, we are 
concerned with the terms which do not depend on $\epsilon R_A$.
Thus when we use the n PN order  integrand,
we 
evaluate $Q_A^{Si_{s+1}i}$ and $R_A^{Ti_{t+1} ij}$. 

To derive $h^{\mu\nu}_{B n=0}$ at the N PN order, we have to know 
$h^{\mu i}_{B n=0}$ up to O($\epsilon^{2N+2}$). Then we have
to use $\mbox{}_{2n+2}\Lambda_N^{\mu\nu}$ to evaluate $Q_A^{Li}$ and
$R_A^{Lij}$ where $n \le N $. Furthermore, since
we can discard the $\epsilon R_A$ dependent terms,  we evaluate
only $Q_A^{Si}$ and $R_A^{Tij}$ where $s = 1,\cdot\cdot\cdot [N-1]$
and $t = 1,\cdot\cdot\cdot [N]$. Thus
at the Newtonian order, $N=0$,  $Q^{Li}$ and $R_A^{Lij}$
do not contribute. At the 1 PN order, $N=1$, we have to evaluate $R_A^{kij}$ 
using $\mbox{}_4[-g t_{LL}^{\mu\nu}]$.  
Up to the 2.5 PN order, we have to evaluate only $Q_A^{i}$,
$Q_A^{ki}$, $R_A^{ij}$, $R_A^{kij}$, and $R_A^{klij}$  
using $\mbox{}_4[-g t_{LL}^{\mu\nu}]$ and
$\mbox{}_6[-g t_{LL}^{\mu\nu}]$.

Explicit calculations required up to the 2.5 PN order are
as follows.

First we calculate
\begin{eqnarray}
\frac{1}{16\pi}\oint_{\pa B_1}
dS_m\mbox{}_4h^{\tau\tau,i}\mbox{}_4h^{\tau\tau,k}y_1^Ly_1^j
&=&
  4(\epsilon R_1)^{l-1}P_1^{\tau 2} C_m^{ikLj} + O((\epsilon R_1)^{l+2}), 
\end{eqnarray}
for l even, 
\begin{eqnarray}
 \mbox{} &&=  O((\epsilon R_1)^{l+1}), 
\end{eqnarray}
for l odd, and  
where 
$$
C^{L} = \frac{1}{4 \pi}\oint d\Omega n^L.
$$

Then, $Q_A^{Li}$ $(l \le 3)$ becomes
\begin{eqnarray}
\epsilon^4 Q_A^{Li} &=&
\epsilon^6\oint_{\pa B_A}dS_k y^{L}_Ay^i_A
\left(\mbox{}_6[-g t_{LL}^{k\tau}] - v_A^k
 \mbox{}_6[-g t_{LL}^{\tau\tau}]\right)   
+ O(\epsilon^8)
\nonumber \\
\mbox{} &=& \frac{1}{2}
 \epsilon^6(\epsilon R_A)^{l-1}P_A^{\tau 2}v_A^k C_k^{Lijj}
 + O(\epsilon^6(\epsilon R_A)^{l+2},\epsilon^8), 
\label{App-EQ-A3}
\end{eqnarray}
for l even, 
\begin{eqnarray}
\mbox{} &&= O(\epsilon^6 (\epsilon R_A)^{l+1}, \epsilon^8), 
\label{App-EQ-A4}
\end{eqnarray}
for l odd.
Thus $Q_A^{Li}$ does not contribute to $h^{\mu\nu}$ up to
the 2.5 PN order. The momentum-velocity relation becomes
$$
P_A^{i} = P_A^{\tau}v_A^i + O(\epsilon^4), 
$$
where we omitted the $\epsilon R_A$ dependence.

As for $R_A^{Lij}$ $(l \le 2)$,  
\begin{eqnarray}
\epsilon^4 R_A^{Lij} &=&
\epsilon^4\oint_{\pa B_A}dS_k y^{L}_Ay^j_A
\mbox{}_4[-g t_{LL}^{ik}]  +
\epsilon^6\oint_{\pa B_A}dS_k y^{L}_Ay^j_A
 \left(\mbox{}_6[-g t_{LL}^{ik}] -
  v_A^k\mbox{}_6[-g t_{LL}^{i\tau}]\right)     
\nonumber \\
\mbox{} &+& 
\epsilon^7\oint_{\pa B_A}dS_k y^{L}_Ay^j_A
\mbox{}_7[-g t_{LL}^{ik}] + O(\epsilon^8)
\nonumber \\
\mbox{} &=& \frac{1}{2}\epsilon^4 (\epsilon R_A)^{l-1}P_A^{\tau 2}
C_k^{Lijk}
+ O(\epsilon^6 (\epsilon R_A)^{l-1},\epsilon^8)
\end{eqnarray}
for l even,
\begin{eqnarray}
&&
\mbox{} =
O(\epsilon^4 (\epsilon R_A)^{l+1}, \epsilon^6 (\epsilon R_A)^{l},
\epsilon^8), 
\end{eqnarray}
for l odd.
Thus $R_A^{Lij}$ also does not contribute to $h^{\mu\nu}$ up to the
2.5 PN order.

Finally, since $Z_A^{Lij}$ contains $J_A^{Lij}$, while $J_A^{Lij}$
contains $Q_A^{Lij}$, when we evaluate $Z_A^{Lij}$, we have to
evaluate not only $R_A^{Lij}$ but also $Q_A^{Lij}$.
From this reason, up to the 2.5
PN order, $Q_A^{kij}$ and $Q_A^{klij}$ appear
(see  Eqs. (\ref{JLiToILi}) and (\ref{ZLijToJLij})). 
From Eqs. (\ref{App-EQ-A3}) and (\ref{App-EQ-A4}),
$Q_A^{kij} = O(\epsilon^4, \epsilon^2(\epsilon R_A))$ and
$Q_A^{klij} = O(\epsilon^2)$.
These results show that they do not contribute to $h^{\mu\nu}$ up to
the 2.5 PN order.

Thus, up to the 2.5 PN order, we have 
\begin{eqnarray}
&&
h^{\tau i}_{B n=0} = 4 \epsilon^4 \sum_{A=1,2}
\left(\frac{P_A^{\tau}v_A^i}{r_A} + \epsilon^2 \frac{M_A^{ki} r^k_A}{2 r_A^3} 
 \right)
 + O(\epsilon^8),
\label{hBtiInAp} \\ 
&&
h^{ij}_{B n=0} = 4 \epsilon^4 \sum_{A=1,2}
\left(\frac{P_A^{\tau}v_A^{i}v_A^j}{r_A}
+ \epsilon^2 \frac{M_A^{k(i}v_A^{j)} r^k_A}{r_A^3} 
 \right)
 + O(\epsilon^8),
 \label{hBijInAp}
\end{eqnarray}
where we used Eqs. (\ref{MomVelRelation}),
(\ref{JijToMij}), (\ref{StrVelRelation}), and 
(\ref{ZkijToAkij})  
and
omitted the $\epsilon R_A$ dependence.

\section{Landau-Lifshitz Pseudo-tensor Expanded With Epsilon}
\label{pNELLPT}

The Landau-Lifshitz pseudo-tensor \cite{LL}
in terms of $h^{\mu\nu}$ which
satisfies the harmonic condition is as follows.
\begin{eqnarray}
(-16 \pi g)t_{LL}^{\mu\nu} &=&
 g_{\alpha\beta}g^{\gamma\delta}
 h^{\mu\alpha}\mbox{}_{,\gamma}
 h^{\nu\beta}\mbox{}_{,\delta}
+ \frac{1}{2}g^{\mu\nu}g_{\alpha\beta}
 h^{\alpha\gamma}\mbox{} _{,\delta}
 h^{\beta\delta}\mbox{}_{,\gamma}
 -2 g_{\alpha\beta}g^{\gamma {\scriptscriptstyle (} \mu}
 h^{\nu {\scriptscriptstyle )}\alpha}\mbox{}_{,\delta}
 h^{\delta\beta}\mbox{}_{,\gamma}
\nonumber \\
\mbox{} &+& 
\frac{1}{2}\left(g^{\mu\alpha}g^{\nu\beta}
			- \frac{1}{2}g^{\mu\nu}g^{\alpha\beta}
		   \right)
           \left(g_{\gamma\delta}g_{\epsilon\zeta}
			- \frac{1}{2}g_{\gamma\epsilon}g_{\delta\zeta}
		   \right)
			h^{\gamma\epsilon}\mbox{} _{,\alpha}
            h^{\delta\zeta}\mbox{}_{,\beta}
\end{eqnarray}

We expand 
the deviation field $h^{\mu\nu}$ in a power series 
of $\epsilon$; 
$$
h^{\mu\nu} = \sum_{n=0}\epsilon^{4+n} \mbox{}_{n+4}h^{\mu\nu}.
$$
Using this equation, we expand $t_{LL}^{\mu\nu}$ with $\epsilon$.
The point to note is that
1) we raise or lower indices with the flat metric $\eta^{\mu\nu}$ and 
$\eta_{\mu\nu}$, 2) $\eta^{\tau\tau}= - \epsilon^2$ and
$\eta_{\tau\tau}= - \epsilon^{-2}$, 3) $\mbox{}_5h^{ij}\mbox{}_{,k} = 0$
and $\mbox{}_7h^{\tau\tau}\mbox{}_{,k} = 0$, 4) $\mbox{}_5h^{\tau\mu} =0$. 
Below, vertical strokes denote that indices between the strokes are 
excluded from (anti-)symmetrization.

\begin{enumerate}
\item $O(\epsilon^4)$
	\begin{eqnarray}
       \mbox{}_4[-16 \pi g t_{LL}^{\tau \tau}] &=& 0,
	 \nonumber \\
 	    \mbox{}_4[-16 \pi g t_{LL}^{\tau i}] &=& 0,
	\nonumber \\	
	  \mbox{}_4[-16 \pi g t_{LL}^{i j}]
	  &=& \frac{1}{4}
	  \left(\mbox{}_4 h^{\tau \tau ,i}
	        \mbox{}_4 h^{\tau \tau ,j}
	  - \frac{1}{2} \delta^{ij}
	        \mbox{}_4 h^{\tau \tau ,k}
	        \mbox{}_4 h^{\tau \tau}_{,k}\right).
\label{tLLij4}
    \end{eqnarray}
 \item $O(\epsilon^6)$
	  \begin{eqnarray}
       \mbox{}_6[-16 \pi g t_{LL}^{\tau \tau}]
	   &=& -\frac{7}{8} \mbox{}_4 h^{\tau \tau ,k}
	        \mbox{}_4 h^{\tau \tau}\mbox{}_{,k},
\label{tLLtt6}	 \\
 	    \mbox{}_6[-16 \pi g t_{LL}^{\tau i}]
	   &=& 2 \mbox{}_4h^{\tau \tau}\mbox{}_{,k}
	       \mbox{}_4h^
		   {\tau {\scriptscriptstyle [}k ,i{\scriptscriptstyle ]}}
		   - \frac{3}{4}\mbox{}_4h^{\tau k}\mbox{}_{,k}
		                \mbox{}_4h^{\tau \tau,i},
\label{tLLti6}		\\	
	  \mbox{}_6[-16 \pi g t_{LL}^{i j}]
	  &=& 4\mbox{}_4h^
	  {\tau {\scriptscriptstyle [}i ,k{\scriptscriptstyle ]}}
	    \mbox{}_4h^{\tau}\mbox{}_{{\scriptscriptstyle [ }k}
		\mbox{}^{,j{\scriptscriptstyle ]}}
		- 2\mbox{}_4h^{\tau \tau ,{\scriptscriptstyle (}i}
		   \mbox{}_4h^{j{\scriptscriptstyle )}k}\mbox{}_{,k}
	\nonumber \\ 
  	   \mbox{} &+& \frac{1}{2}
		\mbox{}_4h^{\tau \tau ,{\scriptscriptstyle (}i}
	       \mbox{}_6h^
		   {|\tau \tau | ,j{\scriptscriptstyle )}}
		+   \frac{1}{2}
		\mbox{}_4h^{\tau \tau ,{\scriptscriptstyle (}i}
	       \mbox{}_4h^
		   {|k}\mbox{}_{k|}\mbox{}^{,j{\scriptscriptstyle )}}
		-  \frac{1}{2}\mbox{}_4h^{\tau \tau}
		\mbox{}_4h^{\tau \tau ,i}\mbox{}_4h^{\tau \tau ,j}   
     \nonumber \\
	   \mbox{} &+& \delta^{ij}
		\left[
					-\frac{3}{8}\mbox{}_4h^{\tau k}\mbox{}_{,k}
					            \mbox{}_4h^{\tau l}\mbox{}_{,l}
                    + \mbox{}_4h^{\tau \tau}\mbox{}_{,k}
					  \mbox{}_4h^{k l}\mbox{}_{,l}
                    + \mbox{}_4h^{\tau k, l}
					  \mbox{}_4h^{\tau}
					  \mbox{}_{{\scriptscriptstyle [}k,
					  l{\scriptscriptstyle ]}} \right.
			\nonumber		  \\ 
        \mbox{} &-& \left. 
					 \frac{1}{4}\mbox{}_4h^{\tau \tau}\mbox{}_{,k}
					            \mbox{}_6h^{\tau \tau,k}
                    -\frac{1}{4}\mbox{}_4h^{\tau \tau,k}
					  \mbox{}_4h^{l}\mbox{}_{l,k}
                    + \frac{1}{4}
					  \mbox{}_4h^{\tau \tau}
					  \mbox{}_4h^{\tau\tau}\mbox{}_{,k}
					  \mbox{}_4h^{\tau\tau,k}
		            \right].
\label{tLLij6}
	  \end{eqnarray}  
 \item $O(\epsilon^7)$
      \begin{eqnarray}
	   \mbox{}_7[-16 \pi g t_{LL}^{\tau \tau}] &=& 0,
	   \label{tLLtt7} \\
	   \mbox{}_7[-16 \pi g t_{LL}^{\tau i}] &=& 0,
	   \label{tLLti7} \\
	   \mbox{}_7[-16 \pi g t_{LL}^{i j}] &=&
	   \frac{1}{2}\left[
	   \mbox{}_4h^{\tau\tau,{\scriptscriptstyle (}i}
	   \mbox{}_5h^{|k}\mbox{}_{k|}\mbox{}^{,j{\scriptscriptstyle )}}
      +\mbox{}_4h^{\tau\tau,{\scriptscriptstyle (}i}
	   \mbox{}_7h^{|\tau\tau|,j{\scriptscriptstyle )}} 
    \right. \nonumber \\
    \mbox{} &-& \left.
    \frac{1}{2}\delta^{ij}\left\{
       \mbox{}_4h^{\tau\tau,l}
	   \mbox{}_5h^{k}\mbox{}_{k,l}
      +\mbox{}_4h^{\tau\tau,k}
	   \mbox{}_7h^{\tau\tau}\mbox{}_{,k}\right\}
	  \right].
	 \label{tLLij7}
      \end{eqnarray}
 \item $O(\epsilon^8)$
 \begin{eqnarray}
	   \mbox{}_8[-16 \pi g t_{LL}^{\tau \tau}] &=& 
       -\frac{7}{4}\mbox{}_4h^{\tau\tau,k}
	   \mbox{}_6h^{\tau\tau}\mbox{}_{,k}
	   \nonumber \\
	   \mbox{} &-&
		\frac{3}{8}\mbox{}_4h^{\tau k}\mbox{}_{,k}
	   \mbox{}_4h^{\tau l}\mbox{}_{,l}
	   +\mbox{}_4h^{\tau k,l}
	   \mbox{}_4h^{\tau}\mbox{}_{{\scriptscriptstyle
	   (}k,l{\scriptscriptstyle )}}
	   \nonumber \\
       \mbox{} &+&
       \frac{1}{4}\mbox{}_4h^{\tau\tau,k}
	   \mbox{}_4h^{l}\mbox{}_{l,k}
	  +\mbox{}_4h^{\tau\tau}\mbox{}_{,k}
	   \mbox{}_4h^{k l}\mbox{}_{,l}
	   \nonumber \\
      \mbox{} &+&
      \frac{7}{8}\mbox{}_4h^{\tau\tau}
	  \mbox{}_4h^{\tau\tau,k}\mbox{}_4h^{\tau\tau}\mbox{}_{,k}, 
\label{tLLtt8}\\
     \mbox{}_8[-16 \pi g t_{LL}^{\tau i}] &=& 
      2 \mbox{}_4h^{\tau}\mbox{}_{k,l}
	    \mbox{}_4h^{k{\scriptscriptstyle[}i,l{\scriptscriptstyle
		]}}
	+ \mbox{}_4h^{\tau i}\mbox{}_{,k}
	   \mbox{}_4h^{kl}\mbox{}_{,l}
	+ \frac{1}{4}\mbox{}_4h^{\tau l}\mbox{}_{,l}
	   \mbox{}_4h^{k}\mbox{}_{k}\mbox{}^{i}
\nonumber \\
    \mbox{} &-& \frac{1}{4}\mbox{}_4h^{\tau\tau,i}
	 \mbox{}_4h^{l}\mbox{}_{l,\tau}
\nonumber \\
     \mbox{} &+&
	  2
	  \mbox{}_6h^{\tau {\scriptscriptstyle[}k,i{\scriptscriptstyle ]}}
	  \mbox{}_4h^{\tau \tau}\mbox{}_{,k}
	  -\frac{3}{4}\mbox{}_6h^{\tau k}\mbox{}_{,k}
	  \mbox{}_4h^{\tau \tau,i}
\nonumber \\
     \mbox{} &+&
	  2 \mbox{}_6h^{\tau \tau}\mbox{}_{,k}
	  \mbox{}_4h^{\tau {\scriptscriptstyle [}k,i{\scriptscriptstyle ]}}
      -\frac{3}{4}\mbox{}_6h^{\tau \tau,i}
	  \mbox{}_4h^{\tau k}\mbox{}_{,k}
\nonumber \\
  \mbox{} &+&
   2\mbox{}_4h^{\tau\tau}
   \mbox{}_4h^{\tau\tau}
   \mbox{}_{,l}\mbox{}_4h^
   {\tau{\scriptscriptstyle[}i,l{\scriptscriptstyle ]}}
 + \frac{3}{4}\mbox{}_4h^{\tau\tau}
   \mbox{}_4h^{\tau\tau, i}
   \mbox{}_4h^
   {\tau l}\mbox{}_{l}
\nonumber \\
    \mbox{} &-& \frac{1}{4}
	 \mbox{}_4h^{\tau k}
	 \mbox{}_4h^{\tau\tau}\mbox{}_{,k}\mbox{}_4h^{\tau\tau,i}
	 + \frac{1}{8}\mbox{}_4h^{\tau i}
	 \mbox{}_4h^{\tau\tau}\mbox{}_{,l}\mbox{}_4h^{\tau\tau,l},
\label{tLLti8}
 \end{eqnarray}
%
%
%
\begin{eqnarray}
\mbox{}_8[-16 \pi g t_{LL}^{i j}]
&=&
 \frac{1}{8}\mbox{}_4h^{ij}
\mbox{}_4h^{\tau\tau,k}\mbox{}_4h^{\tau\tau}\mbox{}_{k}
\nonumber \\
\mbox{} &-&
 2\mbox{}_6h^{\tau\tau,{\scriptscriptstyle (}i}
 \mbox{}_4h^{j{\scriptscriptstyle )}l}\mbox{}_{,l}
+\frac{1}{2}\mbox{}_6h^{\tau\tau,{\scriptscriptstyle (}i}
 \mbox{}_4h^{|l}\mbox{}_{l|}\mbox{}^{,j{\scriptscriptstyle )}} 
- 2\mbox{}_4h^{\tau\tau,{\scriptscriptstyle (}i}
 \mbox{}_6h^{j{\scriptscriptstyle )}l}\mbox{}_{,l}
+\frac{1}{2}\mbox{}_4h^{\tau\tau,{\scriptscriptstyle (}i}
 \mbox{}_6h^{|l}\mbox{}_{l|}\mbox{}^{,j{\scriptscriptstyle )}} 
\nonumber \\
\mbox{} &+&
 2\mbox{}_6h^{\tau l,{\scriptscriptstyle (}i}
  \mbox{}_4h^{j{\scriptscriptstyle )}\tau}\mbox{}_{,l}
-2\mbox{}_6h^{\tau {\scriptscriptstyle (}i,|l|}
  \mbox{}_4h^{j{\scriptscriptstyle )}\tau}\mbox{}_{,l}
+ 2\mbox{}_4h^{\tau l,{\scriptscriptstyle (}i}
  \mbox{}_6h^{j{\scriptscriptstyle )}\tau}\mbox{}_{,l}
-2\mbox{}_6h^{\tau}\mbox{}_{l}\mbox{}^{,{\scriptscriptstyle (}i}
  \mbox{}_4h^{|\tau l|,j{\scriptscriptstyle )}}
\nonumber \\
\mbox{} &-&
 2\mbox{}_4h^{\tau l,{\scriptscriptstyle (}i}
  \mbox{}_4h^{j{\scriptscriptstyle )}}\mbox{}_{l,\tau}
\nonumber \\
\mbox{} &+&
 \mbox{}_4h^{i k,l}\mbox{}_4h^{j}\mbox{}_{k,l}
+\mbox{}_4h^{i k}\mbox{}_{,k}\mbox{}_4h^{j l}\mbox{}_{,l} 
-2\mbox{}_4h^{k l,{\scriptscriptstyle (}i}
  \mbox{}_4h^{j{\scriptscriptstyle )}}\mbox{}_{k,l}
\nonumber \\
\mbox{} &+& 
\frac{1}{2}\mbox{}_4h^{k l,i}\mbox{}_4h_{k l}\mbox{}^{,j}
-\frac{1}{4}\mbox{}_4h^{k}\mbox{}_{k}\mbox{}^{i}
            \mbox{}_4h^{l}\mbox{}_{l}\mbox{}^{j}
\nonumber \\
\mbox{} &+&
 \frac{1}{2}\mbox{}_4h^{\tau\tau,{\scriptscriptstyle (}i}
            \mbox{}_8h^{|\tau\tau|,j{\scriptscriptstyle )}}
\nonumber \\
\mbox{} &+&
 \frac{1}{4}\mbox{}_6h^{\tau\tau,i}
            \mbox{}_6h^{\tau\tau,j}
\nonumber \\			
\mbox{} &+&
 \mbox{}_4h^{\tau\tau}\mbox{}_4h^{\tau i}\mbox{}_{,l}
 \mbox{}_4h^{\tau j,l}
+\mbox{}_4h^{\tau\tau}\mbox{}_4h^{\tau l,i}
 \mbox{}_4h^{\tau}\mbox{}_{l}\mbox{}^{,j}
\nonumber \\
\mbox{} &+& 
 2\mbox{}_4h^{\tau\tau}
  \mbox{}_4h^{\tau\tau,{\scriptscriptstyle (}i} 
  \mbox{}_4h^{j{\scriptscriptstyle )} k}\mbox{}_{,k}
-2\mbox{}_4h^{\tau\tau}
  \mbox{}_4h^{\tau k,{\scriptscriptstyle (}i} 
  \mbox{}_4h^{j{\scriptscriptstyle )} \tau}\mbox{}_{,k}
\nonumber \\
\mbox{} &+&
 \mbox{}_4h^{\tau}\mbox{}_{k}
  \mbox{}_4h^{\tau k,{\scriptscriptstyle (}i} 
  \mbox{}_4h^{|\tau \tau|,j{\scriptscriptstyle )}}
+ \frac{1}{2}
  \mbox{}_4h^{\tau k}\mbox{}_{,k}
  \mbox{}_4h^{\tau \tau,{\scriptscriptstyle (}i} 
  \mbox{}_4h^{j{\scriptscriptstyle )}\tau}
\nonumber \\
\mbox{} &-&
  \frac{1}{2}
  \mbox{}_4h^{\tau \tau}\mbox{}_{,k}
  \mbox{}_4h^{\tau \tau,{\scriptscriptstyle (}i} 
  \mbox{}_4h^{j{\scriptscriptstyle )} k}
- \frac{1}{2}
  \mbox{}_4h^{\tau \tau}
  \mbox{}_4h^{k}\mbox{}_{k}\mbox{}^{,{\scriptscriptstyle (}i} 
  \mbox{}_4h^{|\tau \tau|,j{\scriptscriptstyle )}}
\nonumber \\
\mbox{} &-&
  \mbox{}_4h^{\tau \tau}
  \mbox{}_4h^{\tau \tau,{\scriptscriptstyle (}i}
  \mbox{}_6h^{|\tau \tau|,j{\scriptscriptstyle )}}
-\frac{1}{2}
  \mbox{}_6h^{\tau \tau}
  \mbox{}_4h^{\tau \tau,i}
  \mbox{}_4h^{\tau \tau,j}
\nonumber \\
\mbox{} &+&
 \frac{3}{4}(\mbox{}_4h^{\tau \tau})^2
 \mbox{}_4h^{\tau \tau,i}
  \mbox{}_4h^{\tau \tau,j}
\nonumber \\
\mbox{} &+&  \delta^{i j}
 \left[
         \frac{3}{8}\mbox{}_4h^{\tau \tau}		 
         \mbox{}_4h^{\tau k}\mbox{}_{,k}
	     \mbox{}_4h^{\tau l}\mbox{}_{,l}
       - \mbox{}_4h^{\tau \tau}
         \mbox{}_4h^{\tau \tau}\mbox{}_{,k}
	     \mbox{}_4h^{k l}\mbox{}_{,l}
	   - \mbox{}_4h^{\tau \tau}
	     \mbox{}_4h^{\tau k,l}
		 \mbox{}_4h^{\tau}
		 \mbox{}_{{\scriptscriptstyle [}k,l{\scriptscriptstyle ]}}
\right. \nonumber \\
\mbox{} &+& \left.
         \frac{1}{4}\mbox{}_4h^{\tau \tau}
		 \mbox{}_4h^{\tau \tau, l}
		 \mbox{}_4h^{k}\mbox{}_{k,l}
        -\frac{1}{2}\mbox{}_4h^{\tau k}
		 \mbox{}_4h^{\tau \tau, l}
		 \mbox{}_4h^{\tau}\mbox{}_{k,l}
\right. \nonumber \\
\mbox{} &-& \left.		 
         \frac{1}{4}\mbox{}_4h^{\tau k}
		 \mbox{}_4h^{\tau \tau}\mbox{}_{,k}
		 \mbox{}_4h^{\tau l}\mbox{}_{,l}
        +\frac{1}{8}\mbox{}_4h^{k l}
		 \mbox{}_4h^{\tau \tau}\mbox{}_{,k}
		 \mbox{}_4h^{\tau \tau}\mbox{}_{,l}
\right. \nonumber \\
\mbox{} &+& \left.		
		 \mbox{}_4h^{\tau}\mbox{}_{k,l}
		 \mbox{}_4h^{k l}\mbox{}^{,\tau}
		-\frac{1}{4}\mbox{}_4h^{\tau k}\mbox{}_{,k}
		\mbox{}_4h^{l}\mbox{}_{l,\tau}
\right. \nonumber \\
\mbox{} &+& \left.
		 \frac{1}{2}\mbox{}_4h^{k l,m}
         \mbox{}_4h_{k m,l}
		-\frac{1}{4}\mbox{}_4h^{k l,m}
         \mbox{}_4h_{k l,m}
	    +\frac{1}{8}\mbox{}_4h^{k}\mbox{}_{k,l}
         \mbox{}_4h^{m}\mbox{}_{m}\mbox{}^{,l}
\right. \nonumber \\
\mbox{} &-& \left.
         \frac{3}{4}
		 \mbox{}_4h^{\tau k}\mbox{}_{,k}
         \mbox{}_6h^{\tau l}\mbox{}_{,l}
        +2\mbox{}_4h^{\tau k,l}
		  \mbox{}_6h^{\tau}
		  \mbox{}_{{\scriptscriptstyle [}k,l{\scriptscriptstyle ]}}
\right. \nonumber \\
\mbox{} &+& \left.
		 \mbox{}_4h^{\tau \tau}\mbox{}_{,l}
         \mbox{}_6h^{l k}\mbox{}_{,k}
		-\frac{1}{4}
		 \mbox{}_4h^{\tau \tau,k} 
         \mbox{}_6h^{l}\mbox{}_{l,k}
\right. \nonumber \\
\mbox{} &+& \left.
		 \mbox{}_6h^{\tau \tau}\mbox{}_{,l}
         \mbox{}_4h^{l k}\mbox{}_{,k}
		-\frac{1}{4}
		 \mbox{}_6h^{\tau \tau,k} 
         \mbox{}_4h^{l}\mbox{}_{l,k}	  
\right. \nonumber \\
\mbox{} &+& \left.
         \frac{1}{4}\mbox{}_6h^{\tau \tau}
		 \mbox{}_4h^{\tau \tau}\mbox{}_{,k}
		 \mbox{}_4h^{\tau \tau}\mbox{}^{,k}
        +\frac{1}{2}\mbox{}_4h^{\tau \tau}
		 \mbox{}_4h^{\tau \tau}\mbox{}_{,k}
		 \mbox{}_6h^{\tau \tau}\mbox{}^{,k}
\right. \nonumber \\
\mbox{} &-& \left. 
         \frac{3}{8}(\mbox{}_4h^{\tau \tau})^2
		 \mbox{}_4h^{\tau \tau,k}
		 \mbox{}_4h^{\tau \tau}\mbox{}_{,k}
\right. \nonumber \\
\mbox{} &-& \left. 
         \frac{1}{4}\mbox{}_4h^{\tau \tau,k}
		 \mbox{}_8h^{\tau \tau}\mbox{}_{,k}
\right. \nonumber \\
\mbox{} &-& \left. 
         \frac{1}{8}\mbox{}_6h^{\tau \tau,k}
		 \mbox{}_6h^{\tau \tau}\mbox{}_{,k} 	 
\right]. 
\label{tLLij8}
\end{eqnarray} 
 \item $O(\epsilon^9)$
	\begin{eqnarray}
       \mbox{}_9[-16 \pi g t_{LL}^{\tau \tau}] &=& 0,
\label{tLLtt9}
\\
 	    \mbox{}_9[-16 \pi g t_{LL}^{\tau i}] &=&
		 2\mbox{}_4h^{\tau\tau}\mbox{}_{,l}\mbox{}_7h^{\tau [l,i]}
		 - \frac{1}{4}
		 \mbox{}_4h^{\tau\tau,i}\mbox{}_5h^{l}\mbox{}_{l,\tau}
		 +\frac{3}{4}
		 \mbox{}_4h^{\tau\tau,i}\mbox{}_7h^{\tau\tau}\mbox{}_{,\tau}, 
\label{tLLti9} \\
	  \mbox{}_9[-16 \pi g t_{LL}^{i j}]
	  &=& \frac{1}{2}\mbox{}_4h^{\tau\tau,(i}\mbox{}_9h^{|\tau\tau|,j)}
	  +\frac{1}{2}\mbox{}_4h^{\tau\tau,(i}
	  \mbox{}_7h^{|l}\mbox{}_{l|}\mbox{}^{,j)}
	  - \frac{1}{2}\mbox{}_7h^{\tau\tau}
	   \mbox{}_4h^{\tau\tau,i}\mbox{}_4h^{\tau\tau,j}
\nonumber \\ 
\mbox{} &+&
	  \frac{1}{8}
	  \mbox{}_4h^{\tau\tau,k}\mbox{}_4h^{\tau\tau}\mbox{}_{,k}\mbox{}_5h^{ij}
	  -\frac{1}{2}
	  \mbox{}_4h^{\tau\tau}\mbox{}_{,k}\mbox{}_4h^{\tau\tau,(j}\mbox{}_5h^{i)k}
\nonumber \\ 
\mbox{} &-&
      2\mbox{}_4h^{\tau k,(i}\mbox{}_5h^{j)}\mbox{}_{k,\tau}
 \nonumber \\ 
\mbox{} &+&
      2\mbox{}_4h^{\tau\tau,(i}\mbox{}_7h^{j)\tau}\mbox{}_{,\tau}
 \nonumber \\ 
\mbox{} &-&
      2\mbox{}_4h^{\tau}\mbox{}_{k}\mbox{}^{,(i}\mbox{}_7h^{|\tau k|,j)}
	  +2\mbox{}_4h^{\tau (i}\mbox{}_{|k|}\mbox{}_7h^{|\tau k|,j)}
 \nonumber \\ 
\mbox{} &-&
	  2\mbox{}_4h^{\tau (i}\mbox{}_{|k|}\mbox{}_7h^{j)\tau,k}
	  +2\mbox{}_4h^{\tau k,(i}\mbox{}_7h^{j)\tau}\mbox{}\mbox{}_{,k}
\nonumber \\ 
\mbox{} &+&
      \delta^{ij}\left[
				  2\mbox{}_4h^{\tau k,l}
				  \mbox{}_7h^{\tau}\mbox{}\mbox{}_{[k,l]}
	 - \mbox{}_4h^{\tau\tau}\mbox{}_{,k}
	  \mbox{}_7h^{\tau k}\mbox{}_{,\tau}
				   +
	              \mbox{}_4h^{\tau}\mbox{}_{k,l}
				  \mbox{}_5h^{kl}\mbox{}_{,\tau}
\right.
 \nonumber \\ 
\mbox{} &+& \left.
        \frac{1}{4}
	  \mbox{}_4h^{\tau\tau}\mbox{}_{,\tau}
	  \mbox{}_5h^{l}\mbox{}_{l,\tau}
 -\frac{3}{4}
	  \mbox{}_4h^{\tau\tau}\mbox{}_{,\tau}
	  \mbox{}_7h^{\tau\tau}\mbox{}_{,\tau}
				   +\frac{1}{8}
             	  \mbox{}_4h^{\tau\tau}\mbox{}_{,k}
	              \mbox{}_4h^{\tau\tau}\mbox{}_{,l}\mbox{}_5h^{kl}
			\right.
\nonumber \\
\mbox{} &+& \left.
 \frac{1}{4}
\mbox{}_4h^{\tau\tau,k}\mbox{}_4h^{\tau\tau}\mbox{}_{,k}
\mbox{}_7h^{\tau\tau}
 -\frac{1}{4}
\mbox{}_4h^{\tau\tau,k}\mbox{}_7h^{l}\mbox{}_{l,k}
 -\frac{1}{4}
 \mbox{}_4h^{\tau\tau,k}\mbox{}_9h^{\tau\tau}\mbox{}_{,k}
				 \right].
\label{tLLij9}
 	\end{eqnarray}
\end{enumerate}

\section{Spin-orbit Coupling Force}
\label{SOcoupling}

It is well known that  definition of a dipole moment of the star,  
which we equate to zero to determine the center of the mass of
the star, affects form of the spin-orbit coupling force
(e.g., \cite{Kidder}).
In the paper I we have chosen $d_A^{i} = 0$, where
$$
D_A^{i} = d_A^{i} + \epsilon^2 M_A^{ij}v_A^j + O(\epsilon^4).
$$
 Instead, in this paper we choose $D^i_A = 0$.
The corresponding   
spin-orbit coupling force is 
\begin{eqnarray*}
\left(m_1 \frac{dv_1^i}{d\tau}\right)_{SO} &=&
-\epsilon^4 \frac{2 V^k}{r_{12}^3} 
\left[ \left(m_1 M_2^{il}+ m_2 M_1^{il} \right) \Delta^{lk} 
+ \left( m_1 M_2^{lk}+m_2 M_1^{lk} \right) \Delta^{li} \right] 
\nonumber \\
\mbox{} 
&+& \epsilon^4 \frac{1}{r_{12}^3} 
  \left[ m_2 M_{1}^{lk} v_1^k - m_1 M_{2}^{lk} v_2^k \right]
  \Delta^{li}, 
\end{eqnarray*}
where $\Delta^{ij} = \delta^{ij}-3 n_{12}^i n_{12}^j$.

 \section{Definition of The Mass}
\label{SubtltyOfMass}

In Sec. \ref{NandFPNEOM},
we defined the mass of the star A as    
\begin{eqnarray}
&& 
m_A = \lim_{\epsilon \to 0}P_A^{\tau}.
\label{DefOfMInAp}
\end{eqnarray} 
We shall explain the meaning of this definition.
Also we shall explain the reason we define the mass as above.

Let us consider the origin of the $\epsilon$ dependence of 
$P_A^{\tau}$. The origin can be classified into three categories: 
1) The effect of the motion of the star A itself
and the gravity of the companion star. 
2) the $\epsilon R_A$ dependence which comes from our
splitting of space time, and  
3) the deviation of the scaling law which we impose on 
$\Lambda_N^{\mu\nu}$ on the initial hypersurface.
We shall explain them respectively.

1) This type of $\epsilon$ dependence of $P^{\tau}_A$ can be
specified by solving the evolution equation for $P_A^{\tau}$.
In fact, for example we can rewrite Eq. (\ref{1PNmass}) as
$$ 
P^{\tau}_1 = m_1 + \epsilon^2\frac{1}{2}m_1v_1^2 +
\epsilon^2\frac{ m_1 m_2}{r_{12}} +
\epsilon^2\frac{2 m_1 m_2}{r_{12}} 
+ O(\epsilon^4).
$$
This is the Newtonian energy (times $\sqrt{-g}$)
of the star 1, if one lowers the upper index $\tau$ of $P_1^{\tau}$
down with Newtonian metric and
multiply it minus sign.
At least up to the 2.5 PN order, 
this interpretation really works (see the appendix \ref{MeaningOfPt}).   
Thus we can specify
the  type 1 $\epsilon$ dependence 
of $P_A^{\tau}$ by solving the
evolution equation for $P_A^{\tau}$ functionally.

In fact, our definition of the mass, Eq. (\ref{DefOfMInAp}),
comes from this observation;
if the star A stayed at rest and there were no companion
star, $P_A^{\tau}$ would be the ADM mass of the star A.

2)Since we defined $P^{\tau}_A$ as the volume integral over the
body zone, it must depend on the size of the body zone, $\epsilon R_A$.
As we have stated in Sec. \ref{Arbitrariness}, we ignore this
$\epsilon R_A$ dependence. But even not doing so, when we let
$\epsilon$ go to zero, this $\epsilon R_A$ dependence disappears.
Let us explain how this occurs.

First, because we assume a non-singular source for the star,
the gravitational field on the body zone boundary must be
smooth. In particular we can use the post-Newtonian expanded
gravitational field near and just inside the body zone boundary.
Thus we can estimate the $\epsilon R_A$ dependence using the
post-Newtonian expanded gravitational field. Then, the most
dangerous term (,i.e., the term which seems to diverge when $\epsilon$
goes to zero) in the integrand of the volume integral
Eq. (\ref{DefOfMomentum}) at the n PN order has a form (see the
appendices \ref{RLijandQLi} and \ref{pNELLPT}) 
$$  
\Lambda_N^{\tau\mu} \sim \epsilon^{2n + 4} h^{n-1} h_{,i}h_{,j} \sim 
\left( \frac{m_A}{r_A} \right)^{n-1} \frac{m_A}{r_A^2}\frac{m_A}{r_A^2},
$$
where we omitted the indices. Then we can estimate the
$\epsilon R_A$ dependence at the n PN order as
$$
P_A^{\mu} \sim \epsilon^{-4}(\epsilon R_A)^3 \epsilon^{2n+4}
\left( \frac{m_A}{\epsilon R_A}\right)^{n-1}
\frac{m_A^2}{(\epsilon R_A)^4} \sim \epsilon^{n}.
$$
An explicit calculation (see paper I) shows that  
up to the 1 PN order (,i.e., $n \le 1$ in the above equation) the
$\epsilon R_A$ dependence takes a form like $\epsilon^2 m_A^2/\epsilon R_A$.
Thus the $\epsilon R_A$ dependence vanishes when $\epsilon$ goes to zero.

3)Though we impose scaling law on $T_N^{\mu\nu}$ as Eqs.
(\ref{ScalingTNtt}), (\ref{ScalingTNti}), and (\ref{ScalingTNij}),
there must be deviation from these scalings in the integrand
$\Lambda_N^{\mu\nu}$. 

One reason of this deviation
is that since the velocity of the spinning motion of the star
is set to be $O(\epsilon)$ (slow rotation), the scaling
law on the initial hypersurface can not be exactly satisfied by
$\Lambda_N^{\mu\nu}$.
Besides this, the possibility of obtaining
a sequence of solutions which has the scaling imposed in this paper
has  not been established.
We only assume the existence of such a sequence.

The deviation may also come from the evolution effect. For
instance, as the separation of the binary gets shorter and shorter,
tidal effect stimulates oscillation of the constituent stars.
Such an oscillation may let the stars not satisfy the scaling law.

If we can construct a sequence of solutions which has the 
scaling law  (Eqs. (\ref{ScalingTNtt}), (\ref{ScalingTNti}), and
(\ref{ScalingTNij})) at least as their leading order, 
and 
when we consider a binary system whose separation is long enough,
then we can incorporate the above three deviation into
our formalism by expanding the density $\rho_A$ in an
$\epsilon$ series 
(Let us use the $\rho_A$ instead of $\Lambda_N^{\mu\nu}$ for
notational simplicity).  
$$
\rho_A = \epsilon^{-4}\mbox{}_{(-4)}\rho_A
+ \epsilon^{-3}\mbox{}_{(-3)}\rho_A +
\cdot\cdot\cdot.
$$
From this expansion we have an expansion of the mass in the
$\epsilon$ series.
$$
m_A = \mbox{}_{(0)}m_A + \epsilon \mbox{}_{(1)}m_A + \cdot\cdot\cdot,  
$$
where
$$
\mbox{}_{(n)} m_A = \lim_{\epsilon \to 0}
\int_{B_A}d^3y\mbox{}_{(n-4)}\rho_A.   
$$
Here we need to take $\epsilon$ zero limit to describe the star A as 
a point-like particle.

Thus the $\epsilon$ zero limit in  Eq. (\ref{DefOfMInAp}) is
taken to 1) remove the effect of the motion of the star A itself
and the effect of the existence of the companion star, 2) remove the
dependence of $m_A$ on the body zone boundary, $\epsilon R_A$, and 3)
achieve the point particle limit, while {\it not} removing
the effect of the deviation of $\Lambda_N^{\mu\nu}$
from the scaling law imposed initially.

\section{The 2.5 PN gravitational Field}
\label{DeviationField}

We list the deviation filed $h^{\mu\nu}$
up to the 2.5 PN order. We have checked the harmonic condition up
to the 2.5 PN order; $\mbox{}_{\le 7}h^{\mu\nu}\mbox{}_{,\nu} = 
O(\epsilon^8)$.
We can transform $h^{\mu\nu}$ into the metric
by the formulas given in the appendix \ref{MEwithE}. The resulting
metric perfectly agrees with the result in \cite{BFP}. 

Here are some notations.
\begin{eqnarray}
f^{(\ln S)} =&& \frac{1}{36}(-r_{1}^2 + 3 r_{1} r_{12} 
+ r_{12}^2 - 3 r_{1}r_{2} + 3 r_{12}r_{2} - r_{2}^2) 
\nonumber \\
\mbox{} +&& 
\frac{1}{12}(r_{1}^2 - r_{12}^2 + r_{2}^2) \ln S,
\\
 f^{(1,-1)} =&&  \frac{1}{18}(-r_{1}^2 - 3 r_{1} r_{12} 
- r_{12}^2 + 3 r_{1}r_{2} + 3 r_{12}r_{2} + r_{2}^2)
\nonumber \\
 \mbox{} +&&
\frac{1}{6}(r_{1}^2 + r_{12}^2 - r_{2}^2) \ln S.
\end{eqnarray}
They satisfy $\Delta f^{(\ln S)} = \ln S$ and $\Delta f^{(1,-1)} = r_1/r_2$.
These two potentials (and other useful inverse Laplacian formulas)
can be found in the appendix of \cite{JS98}.

\begin{eqnarray}
h^{\tau\tau} &=& 4\epsilon^4 \sum_{A=1,2} \frac{P_A^{\tau}}{r_A}
\nonumber \\
\mbox{} &&+ \epsilon^6
 \left[7\sum_{a,b=1,2}\frac{P_a^{\tau}P_b^{\tau}}{r_a r_b}
  - \frac{14 P_1^{\tau}P_2^{\tau}}{r_{12}}\sum_{A=1,2}\frac{1}{r_{A}}
  + 2 \frac{\pa^2}{\pa \tau^2}\left(\sum_{A=1,2}P_A^{\tau}r_A\right)
 \right]
 \nonumber \\
\mbox{} &&- \frac{4}{3}\epsilon^7
 \frac{d}{d \tau}\left(\sum_{A=1,2}P_A^{\tau}v_A^2 -
					  \frac{P_1^{\tau}P_2^{\tau}}{r_{12}} \right)
\nonumber \\ 
\mbox{} &&+
\epsilon^8 \left( 8\sum_{A=1,2}\frac{P_A^{\tau 3}}{r_A^3} +
		   \sum_{A=1,2}\frac{P_A^{\tau 2}}{r_A^2}
		   \left(8(\vec n_A\cdot\vec v_A)^2 - 7v_A^2\right)\right)
\nonumber \\
\mbox{} &&+  \epsilon^8 \mbox{}_{8}H^{\tau\tau}
 \nonumber \\ 
\mbox{} &&+
14\epsilon^8\frac{\pa^2}{\pa \tau^2}
\left[- \frac{P_1^{\tau}P_2^{\tau}(r_1+r_2)}{2r_{12}}
\right.
\nonumber \\
\mbox{} &&+ \left.
  \frac{1}{2} \sum_{A=1,2}P_A^{\tau 2}
  \ln \left(\frac{r_A}{{\cal R}/\epsilon}\right) +
P_1^{\tau} P_2^{\tau} \ln \left(\frac{S}{2 {\cal R}/\epsilon}\right)
\right]
 \nonumber \\
\mbox{} &&+ 
\frac{1}{6}\epsilon^8\frac{\pa^4}{\pa \tau^4}
\left(\sum_{A=1,2}P_A^{\tau} r_A^3\right)
\nonumber \\
\mbox{} &&+ \epsilon^9
 \left[4\mbox{}^{(3)}\hat{I}_{orb}^{ij}
   \sum_{A=1,2}\frac{m_A}{r_A}n_A^in_A^j
 - \frac{1}{30}\frac{\pa^5}{\pa \tau^5}
 \left(\sum_{A=1,2}m_Ar_A^4\right)
 \right.
\nonumber \\ 
\mbox{}  &&- \left.
			  \frac{4}{3}\frac{d}{d \tau}
			  \left\{\frac{m_1m_2}{2 r_{12}}
			   \sum_{A=1,2}\left(v_A^2 -
							3(\vec{n}_{12}\cdot \vec{v}_A)^2\right)
				+ \frac{5 m_1m_2
				(m_1+m_2)}{2 r_{12}^2}
\right. \right.
\nonumber \\ 
\mbox{}  &&+ \left. \left.
				 \frac{5 m_1m_2}{2 r_{12}}
				(\vec{n}_{12}\cdot\vec{v}_1)
				(\vec{n}_{12}\cdot\vec{v}_2)
				- \frac{13 m_1m_2}{2 r_{12}}
				(\vec{v}_1\cdot\vec{v}_2)
\right\}
 \right]
+ O(\epsilon^{10}).
\end{eqnarray}
Here 
\begin{eqnarray}
\mbox{}_8 H^{\tau\tau} &=&
P_1^{\tau 2} P_2^{\tau} \left(\frac{8}{r_1^3} + \frac{3}{r_{12}^3} + 
  \frac{173}{3r_1r_{12}^2} - 
  \frac{52}{r_1^2r_{12}} +
  \frac{1}{r_2^{3}} - 
  \frac{r_1^2}{2r_{12}^2r_2^3} - 
  \frac{r_{12}^2}{2r_1^2r_2^3} + 
  \frac{21}{r_1^2r_2}
\right.
\nonumber \\
\mbox{} &&+ \left.
  \frac{9r_1}{2r_{12}^3r_2} + 
  \frac{80}{3r_{12}^2r_2} - 
  \frac{103}{2r_1r_{12}r_2} - 
  \frac{23r_2}{2r_1r_{12}^3} + 
  \frac{15r_2}{2r_1^2r_{12}^2} - 
  \frac{8r_2}{r_1^3r_{12}} - 
  \frac{4r_2^2}{r_1^2r_{12}^3} - 
  \frac{8r_2^2}{r_1^3r_{12}^2} + 
  \frac{8r_2^3}{r_1^3r_{12}^3}\right)
\nonumber \\
\mbox{} &&+
 P_1^{\tau}P_2^{\tau}v_1^2\left(\frac{-10}{3r_1r_{12}} + 
  \frac{5}{r_1r_2} - 
  \frac{31}{3r_{12}r_2} + 
  \frac{2}{r_1S} + \frac{2}{r_{12}S} + 
  \frac{16}{r_2S}\right)
  \nonumber \\
\mbox{} &&+
 P_1^{\tau}P_2^{\tau}\left(\left( \frac{-7}{r_1r_2} - \frac{2}{S^2} - 
     \frac{2}{r_1S} \right) 
(\vec n_1\cdot\vec v_1)^2
- \frac{12(\vec n_1\cdot\vec v_1)(\vec n_{12}\cdot\vec v_1)}{S^2}  
\right.
\nonumber \\
\mbox{} &&+ \left.
  \left( \frac{7}{r_{12}r_2} - \frac{2}{S^2} - 
     \frac{2}{r_{12}S} \right)
   (\vec n_{12}\cdot\vec v_1)^2
\right.
\nonumber \\
\mbox{} &&- \left.
  \frac{16(\vec n_1\cdot\vec v_1)(\vec n_{2}\cdot\vec v_1)}{S^2} - 
  \frac{16(\vec n_{12}\cdot\vec v_1)(\vec n_{2}\cdot\vec v_1)}{S^2} + 
  \left( \frac{-16}{S^2} - \frac{16}{r_2S} \right) 
(\vec n_2\cdot\vec v_1)^2
\right.
 \nonumber \\
 \mbox{} &&+ \left. 
  \left( \frac{28}{3r_1r_{12}} - 
     \frac{4}{r_1r_2} + 
     \frac{28}{3r_{12}r_2} - 
     \frac{16}{r_1S} + \frac{2}{r_{12}S} - 
     \frac{16}{r_2S} \right) 
(\vec v_1\cdot\vec v_2)
+
  \frac{16(\vec n_1\cdot\vec v_2)(\vec n_{12}\cdot\vec v_1)}{S^2}  
\right.
 \nonumber \\
 \mbox{} &&+ \left. 
  \frac{24(\vec n_1\cdot\vec v_2)(\vec n_{2}\cdot\vec v_1)}{S^2}
  + \left( \frac{32}{S^2} + \frac{32}{r_1S} \right) 
(\vec n_1\cdot\vec v_1)(\vec n_{1}\cdot\vec v_2)
-
  \frac{12(\vec n_1\cdot\vec v_1)(\vec n_{12}\cdot\vec v_2)}{S^2}
\right.
 \nonumber \\
 \mbox{} &&+ \left. 	 
  \left( \frac{-2}{S^2} - \frac{2}{r_{12}S} \right) 
(\vec n_{12}\cdot\vec v_1)(\vec n_{12}\cdot\vec v_2)
+\frac{10(\vec n_1\cdot\vec v_1)(\vec n_2\cdot\vec v_2)}{S^2}
			  \right)
+ (1 \leftrightarrow 2)
\end{eqnarray} 
\begin{eqnarray}
h^{\tau i} =&& 4\epsilon^4 \sum_{A=1,2} \frac{P_A^{\tau}v_A^i}{r_A}
\nonumber \\
\mbox{} &&+ \epsilon^6
 \left[
  \sum_{A=1,2}\frac{P_A^{\tau 2}}{r_A^2}
  \left\{(\vec{n}_A\cdot\vec{v}_A)n_A^i + 7 v_A^i
  \right\}
   + \frac{4 P_1^{\tau}P_2^{\tau}}{S r_{12}} (v_{1}^k + v_{2}^k)
   \left(\delta^{ki}- n_{12}^i n_{12}^k\right)
 \right.
\nonumber \\ 
\mbox{}  &&+ \left. 8 P_1^{\tau}P_2^{\tau}(v_1^i + v_2^i)
			  \left(
			   -\frac{1}{r_{12}}\sum_{A=1,2}\frac{1}{r_A}
			   + \frac{1}{r_1 r_2}
              \right)
 \right.
\nonumber \\ 
\mbox{}  &&- \left.			   
			   16 \frac{P_1^{\tau}P_2^{\tau}}{S^2}
			   \left\{
				v_1^k(n_{12}^i - n_1^i)(n_{12}^k+n_2^k)
               +v_2^k(n_{12}^k - n_1^k)(n_{12}^i+n_2^i)
             \right\}
 \right.
\nonumber \\ 
\mbox{}  &&+ \left.
			  12 \frac{P_1^{\tau}P_2^{\tau}}{S^2}
			   \left\{
				v_1^k(n_{12}^k - n_1^k)(n_{12}^i+n_2^i)
               +v_2^k(n_{12}^i - n_1^i)(n_{12}^k+n_2^k)
             \right\}
 \right.
\nonumber \\ 
\mbox{}  &&+ \left.
			  2\sum_{A=1,2}\frac{\pa^2}{\pa \tau^2}
			  \left(P_A^{\tau}v_A^i r_A\right)
			 \right] 
\nonumber \\ 
\mbox{}  &&- 
\frac{2}{3}\epsilon^7\frac{\pa^3}{\pa \tau^3}
\left(\sum_{A=1,2}m_Av_A^ir_A^2 \right)			  
+ O(\epsilon^8).
\end{eqnarray}
\begin{eqnarray}
h^{ij} &=& 4\epsilon^4 \sum_{A=1,2} \frac{P_A^{\tau}v^{i}_Av_A^j}{r_A}
\nonumber \\ 
\mbox{}  &&+ \epsilon^4
 \left[
  \sum_{A=1,2}\frac{P_A^{\tau 2}}{r_A^2}n_A^i n_A^j
  - \frac{8P_1^{\tau}P_2^{\tau}}{r_{12}S}n_{12}^in_{12}^j
  - \frac{8 P_1^{\tau}P_2^{\tau}}{S^2}
  \left(\delta^i\mbox{}_k\delta^j\mbox{}_l
   - \frac{1}{2}\delta^{ij}\delta_{kl}\right)
   \left(
   \vec{n}_{12}-\vec{n}_1 \right)^{(k}
   (\vec{n}_{12} + \vec{n}_2)^{l)}
 \right]
\nonumber \\ 
\mbox{}  &&-
 2\epsilon^5\frac{d^3}{d \tau^3}
 \left(\sum_{A=1,2}P_A^{\tau}z_A^iz_A^j\right)
 \nonumber \\ 
\mbox{}  &&+ \epsilon^6\sum_{A=1,2}
\frac{4 P_A^{\tau 2}}{3r_A^2}
    \left\{
\delta^{ij}
       \left( \frac{-v_A^2}{4} + 
         \frac{(\vec n_A \cdot \vec v_A)^2}{2} \right) 
\right.
\nonumber \\
\mbox{} &&- \left. 
 v_A^2n_A^in_A^j  - 
      \frac{(\vec n_A\cdot\vec v_A)^2n_A^in_A^j}
		 {2} + 
      \frac{(\vec n_A\cdot\vec v_A)
         n_A^{(i}
         v_A^{j)}}{2}+
      \frac{11v_A^i
         v_A^j}{2} \right\} 
 \nonumber \\ 
\mbox{}  &&+ \epsilon^6 \mbox{}_6H^{ij}
\nonumber \\ 
\mbox{} &&+
2\epsilon^6\frac{\pa^2}{\pa \tau^2}
\left[\sum_{A=1,2}P_A^{\tau}v_A^i v_A^jr_A
+\frac{1}{36}\sum_{A=1,2}P_A^{\tau 2}(\delta^{ij} - 3 n_A^i n_A^j)
\right.
\nonumber \\
\mbox{} &&+ \left.
  \frac{\delta^{ij}}{6} \sum_{A=1,2}P_A^{\tau 2}
  \ln \left(\frac{r_A}{{\cal R}/\epsilon}\right) +
\frac{\delta^{ij}}{3}P_1^{\tau} P_2^{\tau}
\ln \left(\frac{S}{2 {\cal R}/\epsilon}\right)
\right.
\nonumber \\
\mbox{} &&- \left.
 4 \left(\delta^{ik}\delta^{jl} - \frac{1}{2}\delta^{ij}\delta^{kl}\right)
P_1^{\tau}P_2^{\tau}
\left(\frac{\pa^2  f^{(\ln S)}}{\pa z^{(k}_1\pa z^{l)}_2}
		 -
		 \frac{\delta_{kl}}{6}\ln S\right)	\right]
\nonumber \\
\mbox{} &&+
\epsilon^7 \frac{\pa^3}{\pa \tau^3}
\left[-\frac{2}{3}\sum_{A=1,2}m_Av_A^{i}v_A^jr_A^2
 +  \frac{ 7 m_1m_2}{r_{12}}\sum_{A=1,2}z_A^iz_A^j  -
\frac{22}{3}m_1m_2r_{12}\delta^{ij} +
\frac{m_1m_2}{3 r_{12}}n_{12}^in_{12}^j\sum_{A=1,2}r_A^2
\right]
\nonumber \\
 \mbox{} &&+  O(\epsilon^8).
\end{eqnarray}
Here $\mbox{}_6H^{ij}$ is
\begin{eqnarray}
 \mbox{}_{6}H^{ij}
&&=4 P^{\tau 2}_1 P_2^{\tau}\left\{
\left( \frac{23}{24r_1^2S} - 
     \frac{3r_1}{4r_{12}^3S} - 
     \frac{163}{24r_{12}^2S} - 
     \frac{11}{12r_1r_{12}S} - 
     \frac{17}{8r_1r_2S} - 
     \frac{r_1^2}
      {8r_{12}^3r_2S} + 
     \frac{17r_1}
      {8r_{12}^2r_2S} + 
     \frac{1}{8r_{12}r_2S}
\right. \right.
\nonumber \\
\mbox{} &&+ \left. \left.
     \frac{11r_2}{12r_{12}^3S}  +
     \frac{17r_2}
      {8r_1r_{12}^2S} + 
     \frac{23r_2}
      {24r_1^2r_{12}S}+
     \frac{11r_2^2}
      {12r_1r_{12}^3S} - 
     \frac{23r_2^2}
      {24r_1^2r_{12}^2S} - 
     \frac{23r_2^3}
      {24r_1^2r_{12}^3S}
     \right)
     \delta^{ij}
\right. 
\nonumber \\
\mbox{} &&+ \left.	 
	 \left( \frac{-1}
      {24r_{12}^3} - 
     \frac{49}{24r_1^2r_{12}} + 
     \frac{r_2^2}
      {24r_1^2r_{12}^3} \right)
	 n_1^i
   n_1^j + 
  \left( \frac{11}
      {3r_1r_{12}^2} - 
     \frac{26}
      {3r_{12}S^2} - 
     \frac{r_2}
      {3r_{12}^2S^2} - 
     \frac{7}{r_{12}^2S}
     \right) n_1^{(i}
     n_{12}^{j)}
\right. 
\nonumber \\
\mbox{} &&+ \left.	   
  \left( \frac{r_1}
      {3r_{12}^2S^2} + 
     \frac{9}{r_{12}S^2} + 
     \frac{2r_2}{3r_{12}^2S^2} + 
     \frac{15}{r_{12}^2S} \right) 
   n_{12}^i
   n_{12}^j
\right. 
\nonumber \\
\mbox{} &&+ \left.	
  \left( \frac{-r_2}
      {3r_{12}^2S^2} + 
     \frac{r_1^2}
      {6r_{12}^3S^2} + 
     \frac{r_1}{3r_{12}^2S^2} - 
     \frac{17}
      {2r_{12}
        S^2} - \frac{r_2^2}
      {6r_{12}^3S^2} \right) 
    n_1^{(i}
   n_2^{j)}
\right. 
\nonumber \\
\mbox{} &&+ \left.	 
  \left( \frac{-r_1}
      {3r_{12}^2S^2} + 
     \frac{25}{3r_{12}S^2}
     \right) n_{12}^{(i}
   n_2^{j)} 
\right\}
\nonumber \\
 \mbox{} &&+ 4 P_1^{\tau}P_2^{\tau}\left\{
-2v_1^2\left(\delta^{ik}\delta^{jl}
				 - \frac{1}{2}\delta^{ij}\delta^{kl}\right)
 \frac{\pa^2 \ln S}{\pa z_1^{(k}\pa z_2^{l)}} +
8\left(\delta^{ik}\delta^{jl}
				 - \frac{1}{2}\delta^{ij}\delta^{kl}\right)
 \frac{\pa^2 \ln S}{\pa z_1^{m}\pa z_2^{(k}}v_{1l)}v_1^m
\right.
\nonumber \\
\mbox{} &&+ \left.
			 4(\vec v_1\cdot\vec v_2)
			 \left(\delta^{ik}\delta^{jl}
				 - \frac{1}{2}\delta^{ij}\delta^{kl}\right)
 \frac{\pa^2 \ln S}{\pa z_1^{k}\pa z_2^{l}}
\right.
\nonumber \\
\mbox{} &&- \left.
 8\delta^{l(i}v_1^{j)} v_2^k \frac{\pa^2 \ln S}{\pa z_1^{k}\pa z_2^{l}}
+4 v_1^iv_2^j \delta^{kl}\frac{\pa^2 \ln S}{\pa z_1^{k}\pa z_2^{l}}
\right.
\nonumber \\
\mbox{} &&+ \left.
	 \delta^{ij}\left(
				  2 \frac{\pa^2 \ln S}{\pa z_1^{k}\pa z_2^{l}}+
				  \frac{3}{2}\frac{\pa^2 \ln S}{\pa z_1^{l}\pa z_2^{k}}
				 \right)v_1^lv_2^k		 
\right.
\nonumber \\
\mbox{} &&- \left.
		 \left(\delta^{ik}\delta^{jl}
				 - \frac{1}{2}\delta^{ij}\delta^{kl}\right)
\frac{\pa^4 f^{(1,-1)}}{\pa z_2^{(k}\pa z_1^{l)}\pa z_1^m \pa z_1^n}v_1^m 
		  v_1^n
\right\}
 \nonumber \\
 \mbox{} &&+ (1 \leftrightarrow 2). 
\end{eqnarray}
Here in the terms $(1 \leftrightarrow 2)$, $f^{(1,-1)}$ is replaced by 
$f^{(-1,1)}$, and this function satisfies $\Delta f^{(-1,1)} =
r_2/r_1$. Its explicit form is the same as Eq. (E2) but with the  
labels 1 and 2 exchanged.

\section{$\chi$ Part}
\label{chipart}

Since $\chi^{\mu\nu\alpha\beta}_N\mbox{}_{,\alpha\beta}$ itself
is conserved, i.e.,
$\chi^{\mu\nu\alpha\beta}_N\mbox{}_{,\alpha\beta\mu} =0$, the
$P_{A \chi}^{\mu}$ by itself is conserved, here we define
 $P_{A\chi}^{\mu}$ as 
\begin{eqnarray}
&&
P_{A \chi}^{\mu} = \epsilon^{-4}\int_{B_A}d^3y
\chi^{\mu\tau\alpha\beta}_N\mbox{}_{,\alpha\beta}.
\end{eqnarray}

First we derive the functional dependence
of $P_{A \chi}^{\mu}$ on  $m_A, v_A^i$, and $r_{12}$. 
By the definition of $\chi^{\mu\nu\alpha\beta}\mbox{}_{,\alpha\beta}$,
$$
16 \pi \chi^{\tau\tau\alpha\beta}\mbox{}_{,\alpha\beta}
= (h^{\tau k}h^{\tau l} - h^{\tau\tau}h^{kl})_{,kl},
$$
$$
16 \pi \chi^{\tau i\alpha\beta}\mbox{}_{,\alpha\beta}
= (h^{\tau \tau}h^{i k} - h^{\tau i}h^{\tau k})_{,\tau k}
+ (h^{\tau k}h^{i l} - h^{\tau i}h^{kl})_{,kl},
$$
thus, we can obtain the functional expressions of $P_{A \chi}^{\mu}$
using Gauss' law. To obtain these expressions up to the 2.5 PN order, 
we need $\mbox{}_4h^{\mu\nu}$ and $\mbox{}_5h^{\mu\nu}$. See
the Sec. \ref{1hPNEOM}. The results are   
\begin{eqnarray}
P_{1 \chi}^{\tau} &=& \epsilon^4\frac{m_1 m_2}{3 r_{12}}
\left[
4 V^2 +\frac{m_2}{r_{12}} - \frac{2 m_1}{r_{12}}
								  \right]
\nonumber \\ 
\mbox{} &-& \epsilon^5
 \frac{2}{3}m_1\mbox{}^{(3)}I_{orb}^l\mbox{}_l
+ O(\epsilon^6), 
\label{Ptchi2hPN} \\
P_{A \chi}^{i} &=& P_{A \chi}^{\tau}v_A^i  + O(\epsilon^6). 
\label{Pichi2hPN}
\end{eqnarray}

Now we move on to the equation of motion for $P_{A \chi}^{\mu}$. 
\begin{eqnarray}
\frac{dP_{A \chi}^{\mu}}{d\tau} &=&
 - \epsilon^{-4} \oint_{\partial B_A}dS_k
 \chi^{\mu k\alpha\beta}_N\mbox{}_{,\alpha\beta}
 + \epsilon^{-4} v_A^k \oint_{\partial B_A}dS_k
 \chi^{\mu \tau\alpha\beta}_N\mbox{}_{,\alpha\beta} . 
\end{eqnarray}
Evaluation of surface integrals gives
\begin{eqnarray}
\frac{dP_{1\chi}^{\tau}}{d\tau} &=&
 - \epsilon^4 \frac{m_1 m_2 (\vec{n}_{12}\cdot\vec{V})}
{3 r_{12}^2}\left[4 V^2
						   + \frac{10 m_2}{ r_{12}}
						   + \frac{4 m_1}{r_{12}}
	 						  \right]
 \nonumber \\
 \mbox{} &-& 
  \epsilon^5\frac{2}{3}m_1\mbox{}^{(4)}I_{orb}^l\mbox{}_l
+ O(\epsilon^6), 
\label{Ptchi2hPNEOM} \\
\frac{dP_{1\chi}^{i}}{d\tau} &=&  - \epsilon^4
 \frac{m_1 m_2 (\vec{n}_{12}\cdot\vec{V})}
{3 r_{12}^2}v_1^i\left[4 V^2
						   + \frac{10 m_2}{ r_{12}}
						   + \frac{4 m_1}{r_{12}}
	 						  \right] 
\nonumber \\ 
\mbox{} &-& \epsilon^4\frac{m_1 m_2^2 }{3 r_{12}^3}n_{12}^i
\left[
4 V^2 +\frac{m_2}{r_{12}} - \frac{2 m_1}{r_{12}}
								  \right]
\nonumber \\ 
\mbox{} &-&
 \frac{2}{3} \epsilon^5\left[m_1v_1^i \mbox{}^{(4)}I_{orb}^l\mbox{}_l  -
					   \frac{m_1m_2 n_{12}^i}{r_{12}^2}
					   \mbox{}^{(3)}I_{orb}^l\mbox{}_l\right].
\label{Pichi2hPNEOM}  
\end{eqnarray}
These two equations, Eqs. (\ref{Ptchi2hPNEOM}) and (\ref{Pichi2hPNEOM})
, are obviously
consistent with Eqs. (\ref{Ptchi2hPN}) and (\ref{Pichi2hPN}).

\section{Metric Expanded With Epsilon}
\label{MEwithE}

Here we give the metric expanded in an $\epsilon$ series for
convenience.

\begin{eqnarray}
\epsilon^{-1}
\sqrt{-g} &=& 1 + \frac{1}{2}\epsilon^2 \mbox{}_4h^{\tau\tau}
\nonumber \\
\mbox{} &+&
 \epsilon^4
 \frac{1}{2}\left[\mbox{}_6h^{\tau\tau}
			-\mbox{}_4h^k\mbox{}_k
			-\frac{1}{4}(\mbox{}_4h^{\tau\tau})^2
			\right]
\nonumber \\
\mbox{} &+&
 \epsilon^5 \frac{1}{2}(\mbox{}_7h^{\tau\tau}
 - \mbox{}_5h^k\mbox{}_k)
\nonumber \\
\mbox{} &+&
 O(\epsilon^6)
\label{ExpandedDetg}
\\
\epsilon^{-1}
g_{\tau \tau} &=& - \epsilon^{-2}
                  + \frac{1}{2}\mbox{}_4h^{\tau\tau}
				  + \epsilon^2\frac{1}{2}
				    \left[\mbox{}_6h^{\tau\tau}
					 + \mbox{}_4h^{k}\mbox{}_{k}
					 -\frac{3}{4}
					 (\mbox{}_4h^{\tau\tau})^2 \right]
\nonumber \\
\mbox{} &+& \epsilon^3 \frac{1}{2}[\mbox{}_7h^{\tau\tau}
                                  +\mbox{}_5h^k\mbox{}_k] 
\nonumber \\
 \mbox{} &+&
\epsilon^4\frac{1}{2}\left[\mbox{}_8h^{\tau\tau}
					 + \mbox{}_6h^k\mbox{}_k
					 -\frac{3}{2}\mbox{}_6h^{\tau\tau}
					 \mbox{}_4h^{\tau\tau}
                     -\frac{1}{2}\mbox{}_4h^k\mbox{}_k
					 \mbox{}_4h^{\tau\tau}
					 + \mbox{}_4h^{\tau k}
					 \mbox{}_4h^{\tau}\mbox{}_k
					 + \frac{5}{8}(\mbox{}_4h^{\tau\tau})^3
					 \right]
 \nonumber \\
 \mbox{}
  &+& \epsilon^5 \frac{1}{2}
\left[\mbox{}_9h^{\tau\tau} + \mbox{}_7h^{k}\mbox{}_k
 - \frac{3}{2}\mbox{}_4h^{\tau\tau}\mbox{}_7h^{\tau\tau}
 - \frac{1}{2}\mbox{}_4h^{\tau\tau}\mbox{}_5h^{k}\mbox{}_k
\right]
 \nonumber \\
 \mbox{}
  &+&
O(\epsilon^6)
\\
\epsilon^{-1}
g_{\tau i} &=& - \epsilon^2 \mbox{}_4h^{\tau i}
- \epsilon^4\left[\mbox{}_6h^{\tau i}
   - \frac{1}{2}\mbox{}_4h^{\tau\tau}\mbox{}_4h^{\tau i}\right]
- \epsilon^5 \mbox{}_7h^{\tau i}
 + O(\epsilon^6)
\\
\epsilon^{-1}
g_{ij} &=& \delta^{ij}
 \left[1 + \epsilon^2 \frac{1}{2}\mbox{}_4h^{\tau \tau} \right] 
\nonumber \\
\mbox{} &+& 
  \epsilon^4
 \left[
  \mbox{}_4h^{ij}
  +\frac{\delta^{ij}}{2}\left(\mbox{}_6h^{\tau\tau}
			   -\mbox{}_4h^k\mbox{}_k
			   -\frac{1}{4}(\mbox{}_4h^{\tau\tau})^2\right)
   \right]
\nonumber \\
\mbox{} &+&
 \epsilon^5\left[\mbox{}_5h^{ij}
			+ \frac{\delta^{ij}}{2}
			\left(\mbox{}_7h^{\tau\tau} -
			\mbox{}_5h^k\mbox{}_k
			\right)\right]
\nonumber \\
 \mbox{} &+&
 O(\epsilon^6)
\end{eqnarray}

\section{Meaning of $P^{\tau}_{A \Theta}$}
\label{MeaningOfPt}

In this section we explain a meaning of  $P_{A \Theta}^{\tau}$.
First of all, we expand in an $\epsilon$ series 
the four velocity of the star A normalized
as $g_{\mu\nu}u_A^{\mu}u_A^{\nu} = -\epsilon^{-2}$, where
$u_A^i = u_A^{\tau}v_A^i$. By the help of the appendix \ref{MEwithE},
we have 
\begin{eqnarray}
u_A^{\tau} &=& 1 +
\epsilon^2\left[\frac{1}{2}v_A^2 + \frac{1}{4}\mbox{}_4h^{\tau\tau}\right]
\nonumber \\
\mbox{} &+&
\epsilon^4\left[
\frac{1}{4}\mbox{}_6h^{\tau\tau} +
\frac{1}{4}\mbox{}_4h^{k}\mbox{}_{k} -
\frac{3}{32}(\mbox{}_4h^{\tau\tau})^2
+ \frac{5}{8}\mbox{}_4h^{\tau\tau}v_A^2
-\mbox{}_4h^{\tau}\mbox{}_{k}v_A^k + \frac{3}{8}v_A^4 \right]
\nonumber \\
\mbox{} &+&
 \epsilon^5\frac{1}{4}\left[\mbox{}_7h^{\tau\tau}
+ \mbox{}_5h^k\mbox{}_k
\right] + O(\epsilon^6).
\label{ExpandedUt} 
\end{eqnarray}
This is a formal series since the metric derived using the point particle
description diverges at the point A.
Now let us regularize this equation by letting
$r_A$ zero while simply discarding the divergent terms.
For example, by this procedure 
$\mbox{}_{\le 6}h^{\tau\tau}$ becomes
(see Eq. (\ref{htt6}))
\begin{eqnarray}
[\mbox{}_{\le 6}h^{\tau\tau}]^{ext}_{1}
&=& 4\epsilon^4 \frac{P^{\tau}_2}{r_{12}}
\nonumber \\
\mbox{} &+& \epsilon^6 \left[
              -2 \frac{m_2}{r_{12}}
			     \{(\vec{n}_{12}\cdot\vec{v}_2)^2-v_2^2\}
                   -16\frac{m_1 m_2}{r_{12}^2}
	+  7 \frac{m_{2}^2}{r_{12}^2}\right],
\label{normarizedhtt6}
\end{eqnarray}
 for the star 1.   In the above equation, $[f]^{ext}_{A}$
means that we regularize  
the quantity f at the star A  
by simply discarding the divergent terms 

Evaluating Eq. (\ref{ExpandedUt}) and (\ref{ExpandedDetg}) by this
procedure, then comparing with Eq. (\ref{2hPNMassEnergy}) with
Eqs. (\ref{DefOfGamma2}) and (\ref{DefOfGamma4}), we find   
at least up to the 2.5 PN order
\begin{eqnarray}
P_{A \Theta}^{\tau} = m_A[\sqrt{-g}u^{\tau}_A]_{A}^{ext}. 
\label{MassEnergyRelationFull}
\end{eqnarray}
It is important that even
after discarding all the divergent terms, there remains non-linear
effect. See Eq. (\ref{DefOfGamma4}) and Eq. (\ref{normarizedhtt6}).  

Eq. (\ref{MassEnergyRelationFull}) is natural. And note that
we have never assumed this relation in advance. This relation has been
derived by solving the evolution equation for
$P_{A \Theta}^{\tau}$ functionally.

The result of the above procedure,  
''discarding all the divergent terms'',   coincides with
the result of some regularization scheme, such as
Hadamard partie finie, at least up to the 2.5 PN order. 
We stress, however, that this coincidence is true for 
the calculation of $P_A^{\tau}$ up to the 2.5 PN order.
This happens  
because only $\mbox{}_4h^{\mu\nu},\mbox{}_5h^{ij},\mbox{}_7h^{\tau\tau},
$ and $\mbox{}_6h^{\tau\tau}$ are needed for the  
calculation of  the 2.5 PN $P_A^{\tau}$ (see Eq. (\ref{ExpandedUt})). When
one calculates the 3 PN $P_A^{\tau}$, for example, 
$\mbox{}_{8}h^{\tau\tau}$ is needed. Then  
''discarding all the divergent terms'' and  Hadamard partie finie 
give different results (to see this, regularize $H_1$ by both ways, 
for example).


\begin{thebibliography}{999}
\bibitem{MTO}A monopole-truncated object is spherically symmetric object 
which is not necessarily compact. Though an equation of motion for such 
objects is useful for describing inspiralling binaries, we stress that 
we have to take into account the effect of multipole moments of stars  
to derive accurate templates especially for binaries in late 
inspiralling phase.
\end{thebibliography}
\end{document}